\documentclass{amsart}
\usepackage{amssymb}
\usepackage{amsmath}
\usepackage{verbatim}

\allowdisplaybreaks

\newcommand{\FF}{\mathcal{F}}
\newcommand{\A}{\mathcal{A}}
\newcommand{\LL}{\mathcal{L}}
\newcommand{\DD}{\mathcal{D}}
\newcommand{\PP}{\mathcal{P}}
\newcommand{\I}{\mathcal{I}}
\newcommand{\HH}{\mathcal{H}}

\newcommand{\RR}{\mathbb{R}}
\newcommand{\CC}{\mathbb{C}}
\newcommand{\ZZ}{\mathbb{Z}}
\newcommand{\bt}{\bullet}
\newcommand{\Fun}{\mathrm{Fun}}
\newcommand{\bulk}{\mathrm{bulk}}
\newcommand{\ora}{\overrightarrow}
\newcommand{\ola}{\overleftarrow}
\newcommand{\dd}{\partial}
\newcommand{\hra}{\hookrightarrow}
\newcommand{\xra}{\xrightarrow}
\newcommand{\be}{\begin{equation}}
\newcommand{\ee}{\end{equation}}

\newcommand{\g}{\mathfrak{g}}
\newcommand{\h}{\mathfrak{h}}
\newcommand{\so}{\mathfrak{so}}
\newcommand{\p}{\mathfrak{p}}
\newcommand{\m}{\mathfrak{m}}
\newcommand{\mr}{\mathrm}
\newcommand{\mc}{\mathcal}

\newcommand{\id}{\mathrm{id}}
\newcommand{\tr}{\mathrm{tr}}
\newcommand{\Str}{\mathrm{Str}}

\newcommand{\ra}{\rightarrow}
\newcommand{\s}{\mathcal{S}}
\newcommand{\End}{\mathrm{End}}
\newcommand{\ad}{\mathrm{ad}}
\newtheorem{theorem}{Theorem}
\newtheorem{proposition}{Proposition}
\newtheorem{lemma}{Lemma}
\newtheorem{corollary}{Corollary}

\theoremstyle{remark}
\newtheorem{Rem}{Remark}

\begin{document}
\title{One-dimensional  Chern-Simons theory}

\author{Anton Alekseev}
\address{Section of Mathematics, University of Geneva, 2-4 rue du Li\`evre, c.p. 64, 1211
Gen\`eve 4, Switzerland}
\email{Anton.Alekseev@unige.ch}

\author{Pavel Mn\"ev}
\address{Petersburg Department of V. A. Steklov Institute of Mathematics, Fontanka 27, 191023 St. Petersburg, Russia}
\address{Institut f\"ur Mathematik, Universit\"at Z\"urich-Irchel,
Winterthurerstrasse 190, CH-8057 Z\"urich, Switzerland}
\email{pmnev@pdmi.ras.ru}

\begin{abstract}
We study a one-dimensional toy version of the Chern-Simons theory.
We construct its simplicial version which comprises features of a low-energy effective gauge theory and of a topological quantum field theory in the sense of Atiyah.
\end{abstract}
\maketitle

\setcounter{tocdepth}{3}\tableofcontents
\tableofcontents

\section{Introduction}
We begin (see section \ref{sec: simplicial CS on circle}) by considering a one-dimensional version of the Chern-Simons theory on a circle. This is a gauge theory in the Batalin-Vilkovisky formalism defined by the action
\be \frac{1}{2}\int (\psi,d\psi)+(\psi,[A,\psi]) , \label{intro 1}\ee
where the field $\psi$ is an odd function on the circle with values in a quadratic Lie algebra $\g$, and the field $A$ is an even 1-form with values in $\g$. We address the problem of constructing an effective BV action induced on a triangulation of the circle.

This problem is interesting by itself since it is related to  discretization of differential geometry. Indeed,  the action (\ref{intro 1}) can be viewed as a generating function for natural operations on differential forms on the circle: the de Rham differential, the wedge product and  the integral over the circle (more precisely, it is a generating function for a unimodular cyclic DGLA structure on $\g$-valued forms, cf. \cite{CM}). In this language, the effective action on a triangulation is a generating function for some discretized (homotopy) version of this structure induced on  cochains of the triangulations (viewed as discrete differential forms).

Another motivation for studying the effective action for (\ref{intro 1})  is that it might give a new insight for constructing  a discrete version of the 3-dimensional Chern-Simons theory. Such a discrete Chern-Simons theory would allow to compute invariants of 3-manifolds as finite-dimensional integrals, and it would be compatible with the gauge symmetry ({\em i.e.} it would satisfy the Batalin-Vilkovisky quantum master equation).

The effective action for the one-dimensional Chern-Simons theory on a triangulated  circle turns out to be given by an explicit but somewhat bizarre formula (\ref{S_Xi explicit}). It immediately raises a number of questions. For instance, the result is expected to satisfy the quantum master equation (QME) and to be compatible with simplicial aggregations (merging several 1-simplices of the triangulation). How can we check this directly? Another desire is to represent the result (\ref{S_Xi explicit}) in a ``simplicially-local'' form.

It turns out that answers to these questions come from the following construction. We give a new definition of the one-dimensional simplicial Chern-Simons theory in the ``operator formalism'', {\em i.e.} in the language of Clifford algebras $Cl(\g)$  (section \ref{sec: operator formalism, 1d simp cs}). The partition function for a simplicial complex is an element of $Cl(\g)^{\otimes i}$ (where $i$ is half the number of boundary points of the simplicial complex), and it given by  a product of local $Cl(\g)$-valued expressions (\ref{Z_I}) for 1-simplices. In particular, for a triangulated circle the partition function takes values in numbers (and it also depends on  simplicial ``bulk fields''). In section \ref{sec: back to path integral, non-abelian case}, we establish the equivalence between the operator formalism and the path integral formalism of section \ref{sec: simplicial CS on circle}. Consistency with simplicial aggregations is checked straightforwardly in the operator language (section \ref{sec: aggregations}). The partition function for an interval can be shown to satisfy equation (\ref{QME for Z_I}):
$$\hbar\, \frac{\dd}{\dd\tilde\psi^a}\,\frac{\dd}{\dd A^a}Z_\I+ \frac{1}{\hbar}\, \left[\frac{1}{6} f^{abc} \hat\psi^a\hat\psi^b\hat\psi^c,Z_\I\right]_{Cl(\g)}=0$$
which is interpreted  as a version of the quantum master equation adjusted for the presence of the boundary. This immediately implies the QME with boundary contributions for arbitrary one-dimensional simplicial complex (\ref{QME for Z_Theta}) and the usual QME for the triangulated circle (\ref{QME for S_Xi}).

To formulate one-dimensional simplicial Chern-Simons theory in the spirit of Atiyah's axioms of TQFT (section \ref{sec: Segal's axioms}), we 
choose a complex polarization of $\g$:
$$\g_\CC=\h\oplus\bar\h$$
(which can always be introduced if $\g$ is even-dimensional; however, by introducing a complex polarization  we break the $O(\g)$-symmetry of the original problem). The Clifford algebra $Cl(\g)$ is isomorphic to the matrix (super-)algebra $\End(\wedge^\bt\h)=\End(\Fun(\Pi\h))$. Therefore, the space of states associated to a point in the one-dimensional Chern-Simons theory is $\HH_{pt}=\Fun(\Pi\h)$ --- the super vector space of polynomials in $(\dim\g)/2$ odd variables. The super-space $\HH_{pt}$ is endowed with an odd third-order differential operator $\delta$. One-dimensional ``cobordisms'' are now equipped with triangulations. To a triangulated cobordism $\Theta$ we associate the ``space of bulk fields'' $\FF^\bulk_\Theta$, equipped with the BV Laplacian $\Delta^\bulk_\Theta$. The partition function for a triangulated cobordism satisfies the quantum master equation (\ref{QME for Z^rho_Theta}). In addition to the operations of gluing and disjoint union (which are standard in Atiyah's picture),  simplicial aggregations are allowed for triangulated cobordisms. The original continuum theory can be thought of as the simplicial theory in the limit of dense triangulation.

Matrix elements of the partition function for a triangulated cobordism can be written as path integrals for the one-dimensional Chern-Simons with BV gauge fixing in the bulk and holomorphic-antiholomorphic boundary conditions (section \ref{sec: back to path integral, non-abelian case}). This brings us back to the formalism of effective BV actions. The action for an interval is given by a Gaussian integral and it is easy to compute it explicitly (\ref{W_I}).



\subsection{Acknowledgements}
We wish to thank Alberto Cattaneo and Andrei Losev for enlightening discussions on the subject. 
Research of A.~A. was supported in part by the grants of the Swiss National
Science Foundation number 200020-129609 and 200020-126817;
P.~M. acknowledges partial support by SNF Grant 200020-121640/1 and by RFBR Grants 08-01-00638, 09-01-12150.

\subsection{Authorship}
The idea of looking at the one-dimensional Chern-Simons theory and
some parts of section \ref{sec 3} (operator formalism approach) are joint work of both authors (the idea to use operator formalism for the one-dimensional Chern-Simons theory was suggested by A. A.). Other parts  of the paper are due to P. M.

\section{Simplicial Chern-Simons theory on the circle}
\label{sec: simplicial CS on circle}

In this section we study the Chern-Simons theory on the circle in  Batalin-Vilkovisky (BV) formalism and construct an effective BV action induced on cochains of a triangulation. Much of this discussion is inspired by  \cite{CM} and \cite{simpBF} . In particular, the reader is referred to sections 2 and 3.2 of \cite{CM} for details of the effective BV action construction.

\subsection{Continuum theory on the circle: fields, BV structure, action}
\label{sec: continuum theory}

Let $\g$ be a quadratic Lie algebra with Lie bracket $[,]$ and non-degenerate ad-invariant pairing $(,)$.
We will denote by $\{T^a\}$ an orthonormal basis in $\g$ and by $f^{abc}=(T^a,[T^b,T^c])$ the structure constants in this basis. We will also use the Einstein summation convention for the Lie algebra indices.

The Chern-Simons theory on a 3-manifold $M$ can be constructed as an AKSZ sigma model \cite{AKSZ} with the space of fields
$$\FF=\mr{Maps}(\Pi TM, \Pi\g) = \Pi\g\otimes \Omega^\bt(M)  . $$
That is, $\FF$ is the space of maps of super-manifolds from the parity-shifted tangent bundle of $M$ to the parity-shifted Lie algebra.
Equivalently, this is  the space of differential forms on $M$ with values in $\Pi\g$.
From the canonical integration measure on $\Pi TM$ and thew even symplectic structure
$\omega_{\Pi\g}=\frac{1}{2} \delta X^a \wedge \delta X^a$
on $\Pi\g$ (we denote by $\{X^a\}$ the set of odd coordinates on $\Pi\g$ associated to the orthonormal basis $\{T^a\}$ on $\g$) one constructs an odd symplectic form (the ``BV 2-form'') on $\FF$:
\be \omega=\frac{1}{2}\int_M (\delta\alpha, \delta\alpha)    . \label{Omega}\ee
Here the superfield $\alpha$ is the canonical odd map (the parity-shifted identity operator)
\be \alpha: \FF \ra \g\otimes \Omega^\bt(M) \label{superfield as parity-shifted id}\ee
which can be viewed as the generating function for coordinates on $\FF$ with values in $\g$-valued differential forms on $M$. By splitting $\alpha$ into components according to the degrees of differential forms we obtain
$$\alpha=A^{(0)}+A^{(1)}+A^{(2)}+A^{(3)}$$
where $A^{(p)}$ takes values in $\g$-valued $p$-forms.\footnote{In the BV formalism, $A^{(1)}$ is the ``classical field'', $A^{(0)}$ is the ``ghost''; $A^{(2)}$ and $A^{(3)}$ are the ``anti-fields'' for $A^{(1)}$ and $A^{(0)}$, respectively.} Since $\alpha$ is totally odd, $A^{(0)}$ and $A^{(2)}$ are intrinsically odd, $A^{(1)}$ and $A^{(3)}$ are intrinsically even (the intrinsic parity is the total parity minus the de Rham degree modulo 2). The Chern-Simons action is built of the 1-form $\frac{1}{2} X^a \wedge \delta X^a$ on $\Pi\g$ (which is a primitive for $\omega_{\Pi\g}$)
and the odd function on $\Pi\g$
\be \theta=\frac{1}{6} f^{abc} X^a X^b X^c \label{theta}\ee
which satisfies $\{\theta,\theta\}_{\Pi\g}=0$. The action is given by formula,
\be S=\int_M \frac{1}{2}(\alpha, d\alpha)+\frac{1}{6} (\alpha,[\alpha,\alpha]) .  \label{AKSZ action}\ee
By the general construction \cite{AKSZ}, $S$ satisfies the classical master equation
$$\{S,S\}=0$$
where $\{,\}$ is the BV anti-bracket on functions on $\FF$ defined by the odd symplectic form $\omega$.

We would like to define the one-dimensional Chern-Simons theory on the circle $\s^1$ by substituting $M=\s^1$ into the construction described above. Then, the space of fields becomes
$$\FF=\mr{Maps}(\Pi T \s^1, \Pi\g)=\Pi \g\otimes\Omega^0(\s^1) \oplus \Pi \g\otimes\Omega^1(\s^1)$$
%
The superfield $\alpha$ can now  be written as
\be \alpha=\psi+A ,  \label{alpha 1D splitting}\ee
where the component $\psi=T^a \psi^a(\tau)$ takes values in $\g$-valued functions on the circle and is intrinsically odd ($\tau$ is the coordinate on $\s^1$); and the component $A=d\tau\, T^a A^a(\tau)$ takes values in $\g$-valued 1-forms on the circle and is intrinsically even. Thus, $\{\psi^a(\tau), A^a(\tau)\}$ are odd and even coordinates on 
$\FF$, respectively.
The space $\FF$ is equipped with an odd symplectic structure (\ref{Omega}):
\be \omega=\int_{\s^1}(\delta\psi, \delta A), \label{BV form on circle}\ee
defining the anti-bracket  $\{\bt,\bt\}: \Fun(\FF)\times \Fun(\FF)\ra \Fun(\FF)$,
$$\{f,g\}=\int_{\s^1} d\tau\; f \left(\frac{\ola \delta}{\delta\psi^a(\tau)}\frac{\ora\delta}{\delta A^a(\tau)}- \frac{\ola \delta}{\delta A^a(\tau)}\frac{\ora\delta}{\delta \psi^a(\tau)}\right) g$$
and the BV Laplacian $\Delta: \Fun(\FF)\ra \Fun(\FF)$
$$\Delta f=\int_{\s^1} d\tau\; \frac{\delta}{\delta \psi^a(\tau)} \frac{\delta}{\delta A^a(\tau)} f .$$
Note that the operator $\Delta$ is ill-defined on local functionals.

The action (\ref{AKSZ action}) can be written in terms of components (\ref{alpha 1D splitting}) of the superfield as
\be S=\frac{1}{2}\int_{\s^1}\left((\psi,d\psi)+(\psi,[A,\psi])\right) . \label{action on circle}\ee
Here $d$ is the de Rham differential on $\s^1$. By the general AKSZ construction\footnote{Or, in the algebraic language, due to relations (Leibniz identity, Jacobi identity, cyclicity of differential, cyclicity of Lie bracket) in the cyclic dg Lie algebra $\g\otimes \Omega^\bt(\s^1)$, cf. \cite{CM}.}, the action $S$ satisfies the classical master equation
$$\{S,S\}=0.$$
Na\"ively, one could also say that the unimodularity of $\g$ implies unimodularity of $\g\otimes\Omega^\bt(\s^1)$, and therefore the quantum master equation is fulfilled
$$\frac{1}{2}\{S,S\}+\hbar \Delta S =0.$$
However, $\Delta S$ is ill-defined in continuum theory.


\begin{Rem}
The $\ZZ_2$-grading on the space of fields of the Chern-Simons theory on a 3-manifold can be promoted to a $\ZZ$-grading (by setting $\FF=\mr{Maps}(T[1]M,\g[1])$) in such a way that the odd symplectic form $\Omega$ attains grade\footnote{Grade is defined as the total degree minus the de Rham degree of a differential form.} $-1$ (so that the anti-bracket has degree $+1$) and the action $S$ is in degree zero. However, this does not apply to the one-dimensional Chern-Simons theory which is essentially $\ZZ_2$-graded: there is no consistent $\ZZ$-grading on the space of fields.
\end{Rem}



\subsection{Effective action on the cohomology of the circle}
\subsubsection{Harmonic gauge.}
\label{sec: coh of circle, gauge}
Let's split the space of differential forms on the circle into constant 0- and 1-forms and those with vanishing integral\footnote{Properties of being constant for a 1-form and being of integral zero for a 0-form are non-covariant. This is not a problem as choosing a gauge always relies on introducing some additional structure. In the case of harmonic gauge, this extra structure is the parametrization of the circle.}:
$\Omega^\bt(\s^1)=\Omega'^\bt(\s^1)\oplus \Omega''^\bt(\s^1)$
where
\begin{eqnarray}
\Omega'^\bt(\s^1) & = & \{f + d\tau\, g \; | \;  f,g\in \RR\} \label{splitting on circle: IR}  , \\
\Omega''^\bt(\s^1) & = & \{f''(\tau) + d\tau\, g''(\tau) \; | \;  \int_{\s^1} d\tau\, f''(\tau)=0,\, \int_{\s^1} d\tau\, g''(\tau)=0\} . \label{splitting on circle: UV}
\end{eqnarray}
It induces the splitting for  fields into infrared and ultraviolet parts $\FF=\FF'\oplus\FF''$, where
\begin{eqnarray*}
\FF' & = & \{\psi_0 + d\tau\, A_0 \; | \;  \psi_0\in \Pi\g, A_0\in \g\} , \\
\FF'' & = & \{\psi''+ A'' \;|\; \int_{\s^1}d\tau\, \psi''(\tau)=0, \, \int_{\s^1}A''=0\} .
\end{eqnarray*}
This splitting 
respects both the BV 2-form and the de Rham differential.
We define the Lagrangian subspace $\LL\subset\FF''$ as
\be \LL=\{\psi''+A'' \;|\; A''=0\} . \label{L}\ee

\subsubsection{Effective action on cohomology.}
\footnote{More precisely, we consider the effective action as a function on $\Pi\g\otimes H^\bt(\s^1)$, {\em i.e.} on the parity-shifted de Rham cohomology of the circle (which we represented by harmonic forms) with coefficients in $\g$.}
We define the effective action $W$ on $\FF'$ by the fiber BV integral
\begin{multline}
e^{\frac{1}{\hbar}W(\psi_0,A_0,\hbar)}=\int_\LL e^{\frac{1}{\hbar}S(\psi_0+\psi'',\, d\tau\,A_0+A'')}=
\\= \int \DD \psi''\; e^{\frac{1}{2\hbar}\int_{\s^1} \left(\psi_0+\psi'',(d+d\tau\,\ad_{A_0})(\psi_0+\psi'')\right)} .
\label{BV induction to cohom}
\end{multline}
The Lagrangian subspace $\LL\subset\FF''$ is  uniquely\footnote{This a special property of the one-dimensional theory related to the fact that the de Rham operator $d: \Omega''^0(\s^1)\ra \Omega''^1(\s^1)$ is an isomorphism, and there is unique chain homotopy $K=d^{-1}: \Omega''^1(\s^1)\ra \Omega''^0(\s^1)$.} fixed by the requirement that the ``free'' part of the action  $\frac{1}{2}\int_{\s^1}(\psi,d\psi)$  be non-degenerate when restricted to $\LL$.

Integral (\ref{BV induction to cohom}) is Gaussian,  and it yields the following result.
\begin{proposition}
The effective BV action $W$ of the  one-dimensional Chern-Simons theory  is given by
\be e^{\frac{1}{\hbar}W(\psi_0,A_0,\hbar)} = {\det}_\g^{1/2}\left(\frac{\sinh\frac{\ad_{A_0}}{2}}{\frac{\ad_{A_0}}{2}}\right)\cdot e^{-\frac{1}{2\hbar} (\psi_0,\ad_{A_0}\psi_0)} \label{W on coh of circle}  .  \ee
\end{proposition}
A simple form of the 0-loop part is due to the fact that multiplication by  constant 1-forms respects the splitting of forms into infrared and ultraviolet parts (\ref{splitting on circle: IR},\ref{splitting on circle: UV}).\footnote{Higher order terms in the 0-loop effective action would correspond to Massey operations on the de Rham cohomology (cf. \cite{CM}, \cite{simpBF}). In the case of the circle, Massey operations vanish.} The functional determinant is easily computed {\em e.g.} by using the exponential basis $\{e^{2\pi i k\tau}\}$ for 0-forms on the circle.

The effective action (\ref{W on coh of circle}) satisfies the quantum master equation
$$\frac{\dd}{\dd\psi_0^a}\frac{\dd}{\dd A_0^a} e^{\frac{1}{\hbar}W(\psi_0,A_0,\hbar)}=0. $$
The classical master equation is implied by the Jacobi identity and by cyclicity property of the Lie bracket on $\g$; the quantum part of master equation follows from the fact that the one-loop part of $W$ is manifestly ad-invariant.

\subsection{Simplicial Chern-Simons action on circle}
\label{sec: simplicial CS on circle 1.3}

Let's assume that the circle $\s^1$ is glued of $n$ intervals $\I_1=[\p_1,\p_2],\ldots,\I_n=[\p_n,\p_1]$ (where $\p_1,\ldots,\p_n$ is a cyclically ordered collection of points on the circle), and that each interval $\I_k$ is equipped with a coordinate function $\tau: \I_k \ra [0,1]$. We will denote a point of $\I_k$ with coordinate $\tau$ by
$(k,\tau)$. We will assume $n$ to be odd (otherwise, our gauge will be inconsistent, see remark \ref{Rem: n odd}). We denote this ``triangulation'' of the circle by $\Xi_n$.

\subsubsection{Cyclic Whitney gauge.}
\label{sec: PL gauge}
Splitting of fields into infrared and ultraviolet parts is defined by splitting for differential forms on the circle $\Omega^\bt(\s^1)=\Omega'^\bt_{\Xi_n}(\s^1)\oplus \Omega''^\bt_{\Xi_n}(\s^1)$, where we split 0-forms into continuous piecewise-linear ones and those with vanishing integrals over each $\I_k$, and we split 1-forms into piecewise-constant ones and the orthogonal complement of piecewise-linear 0-forms:
\begin{multline}
\Omega'^\bt_{\Xi_n}(\s^1)=\{f'+d\tau\, g' \quad |\quad f'|_{\I_k}=(1-\tau) f_k + \tau f_{k+1},\, g'|_{\I_k}=g_k \; \forall k \} \label{PL splitting for Omega}\\
\Omega''^\bt_{\Xi_n}(\s^1)=\{f''+d\tau\, g'' \quad | \quad \int_{\I_k} d\tau f''=0,\, \int_{I_k} \tau d\tau\,  g''+ \int_{I_{k+1}}  (1-\tau) d\tau\, g''=0 \; \forall k \}
\end{multline}
where $f_k,g_k\in\RR$ are numbers. Thus, $\Omega'^\bt_{\Xi_n}(\s^1)\cong \RR^{n|n}$ (as a super-space). As a cochain complex, $\Omega'^\bt_{\Xi_n}(\s^1)$ is isomorphic to $C^\bt(\Xi_n)$, the cochain complex of the simplicial complex $\Xi_n$.  As in section \ref{sec: coh of circle, gauge}, this splitting  agrees with the de Rham differential, and  the associated splitting for fields $\FF=\FF'\oplus \FF''$ agrees with the BV 2-form.

We call splitting (\ref{PL splitting for Omega}) the ``cyclic Whitney gauge'', because our representatives $\Omega'^\bt$ for cell cochains of triangulation are exactly the Whitney forms \cite{Whitney} for $\Xi_n$. The word ``cyclic'' indicates that  $\Omega''^\bt$ is constructed as an orthogonal complement of $\Omega'^\bt$ with respect to the Poincar\'e pairing $\int_{\s^1}\bt\wedge\bt$.

Coordinates on $\FF'$ are given by values of $\psi'$ at the vertices of triangulation:
\be \psi_k=\psi'(\p_k)\in \Pi\g \label{psi_k} , \ee
and by integrals of $A$ over intervals:
\be A_k=\int_{\I_k} A'\in \g \label{A_k} . \ee
The BV 2-form on $\FF'$ is given by
\be\omega'=\sum_{k=1}^n \frac{\delta \psi_k^a+\delta \psi_{k+1}^a}{2}\wedge \delta A_k^a \label{simplicial omega} . \ee
We will denote $\FF'=\FF_{\Xi_n}$ to emphasize its dependence on $n$. We have $\FF_{\Xi_n}\cong \Pi\g\otimes C^\bt(\Xi_n)$.

\begin{Rem} \label{Rem: n odd}
The requirement that $n$ be odd is needed since for $n$ even  the piecewise-linear 0-form
$$f(k,\tau)=(-1)^k (\tau-1/2)$$
belongs to both the infrared and ultraviolet subspaces.
\end{Rem}

As in section \ref{sec: coh of circle, gauge}, we define the Lagrangian subspace $\LL\subset\FF''$ by setting the 1-form part of the ultraviolet field to zero (\ref{L}).

\subsubsection{Chain homotopy, dressed chain homotopy.}
\label{sec: chain homotopy}
Let us define the infrared projector $\PP': \Omega^\bt(\s^1)\ra \Omega'^\bt_{\Xi_n}(\s^1)$ by formula
\begin{multline}
f+d\tau\cdot g \mapsto \\ \mapsto\sum_{k=1}^n \left(\int_{\I_k}d\tau'\, f-(1-2\tau) \cdot\left(\int_{\I_{k+1}}d\tau'\, f-\int_{\I_{k+2}}d\tau'\, f+\cdots+\int_{\I_{k-2}}d\tau'\, f-\int_{\I_{k-1}}d\tau'\, f\right)\right)\theta_{\I_k}+\\
+d\tau\sum_{k=1}^n \left(\int_{\I_k}d\tau'\, g+\int_{\I_{k+1}}(1-2\tau') d\tau'\,g-\int_{\I_{k+2}}(1-2\tau') d\tau'\,g+\cdots-\int_{\I_{k-1}}(1-2\tau') d\tau'\,g \right)\theta_{\I_k}
\label{P'} ,
\end{multline}
where $\theta_{\I_k}$ is the function on the circle with value $1$ on $\I_k$ and zero elsewhere.

The chain homotopy $\kappa:\Omega^1(\s^1)\ra \Omega^0(\s^1)$ is uniquely defined  by the properties
\begin{eqnarray}
d\,\kappa+\kappa\, d & = & \id-\PP' \label{K 1} , \\
\PP'\kappa & = & 0 \label{K 2} , \\
\kappa \PP' & = & 0 \label{K 3} .
\end{eqnarray}
\begin{lemma}
The operator $\kappa$ defined by relations (\ref{K 1}), (\ref{K 2}), (\ref{K 3}) acts on the 1-form $d\tau\cdot g\in \Omega^1(\s^1)$ by\footnote{Recall that $(k,\tau)$ denotes a point on the circle which belongs to the interval $\I_k$ and which has a local coordinate $\tau$. Hence, the integral kernel here is actually a function on $\s^1\times \s^1$.}
\be \kappa(d\tau\cdot g)(k,\tau)=\sum_{k'=1}^n\int_{\I_{k'}}d\tau' \kappa((k,\tau),(k',\tau'))\, g(k',\tau') , \label{K via integral kernel}\ee
where the integral kernel is given by
\be \kappa((k,\tau),(k',\tau'))=\left\{\begin{array}{ll}
\theta(\tau-\tau')-\frac{1}{2}-\tau+\tau' &  \mbox{if}\; k=k' \\
(-1)^{k-k'} 2 (\frac{1}{2}-\tau) (\frac{1}{2}-\tau') & \mbox{if}\; k'<k<k'+n
\end{array}\right. \label{kappa}\ee
and $\theta$ is the unit step function.
In addition, the kernel has the anti-symmetry property:
$$\kappa((k',\tau'),(k,\tau))=-\kappa((k,\tau),(k',\tau'))$$
\end{lemma}
To obtain formula (\ref{kappa}), one observes that relations (\ref{K 1}) and  (\ref{K 2}) imply the differential equation
\be \frac{\dd}{\dd \tau} \kappa((k,\tau),(k',\tau'))= \delta_{k,k'} \delta(\tau-\tau')+ C_k(k',\tau') \label{kappa 1}\ee
subject to conditions
\be \int_0^1  d\tau\, \kappa((k,\tau),(k',\tau'))=0,\qquad \kappa((k,1),(k',\tau'))=\kappa((k+1,0),(k',\tau')) \qquad \forall k . \label{kappa 2}\ee
Here $C_k(k',\tau')$ are some functions independent of $\tau$. Solving (\ref{kappa 1}) together with (\ref{kappa 2}) immediately yields (\ref{kappa}). This proves the uniqueness property. In order to prove existence, one checks that (\ref{kappa}) satisfies (\ref{K 1}), (\ref{K 2}), (\ref{K 3}).

We will also need the ``dressed'' chain homotopy ({\em i.e.} dressed by the connection) $\kappa_{A'}: \g\otimes\Omega^1(\s^1)\ra \g\otimes\Omega^0(\s^1)$, where $A'=\sum_{k=1}^n A_k\theta_{\I_k}$ is a piecewise-constant $\g$-valued 1-form on the circle. The operator $\kappa_{A'}$ is uniquely defined by the properties
\begin{eqnarray}
(\id-\PP') d_{A'}\,\kappa_{A'}+\kappa_{A'}\, d_{A'} (\id-\PP') & = & \id-\PP' \label{K_A prop 1} , \\
\PP'\kappa_{A'} & = & 0 \label{K_A prop 2} , \\
\kappa_{A'} \PP' & = & 0  \label{K_A prop 3} ,
\end{eqnarray}
where $d_{A'}=d+\ad_{A'}$. One can summarize these properties by saying that $\kappa_{A'}: \Omega''^1_{\Xi_n}(\s^1)\ra \Omega''^0_{\Xi_n}(\s^1)$ is the inverse of the operator $(\id-\PP')d_{A'}(\id-\PP'):\Omega''^0_{\Xi_n}(\s^1)\ra \Omega''^1_{\Xi_n}(\s^1)$. The dressed chain homotopy can either be expressed perturbatively as a series
$$\kappa_{A'}=\kappa-\kappa\,\ad_{A'}\,\kappa+\kappa\,\ad_{A'}\,\kappa\,\ad_{A'}\,\kappa-\cdots ,$$
or it can be computed  explicitly by solving the differential equation (\ref{K_A prop 1}).

To present an explicit formula for the integral kernel of $\kappa_{A'}$ (defined as in (\ref{K via integral kernel})), we have to first  introduce some notation. Let $F(\A,\tau)$  be an  $\End(\g)$-valued function on the interval $[0,1]$ depending on an anti-symmetric matrix $\A\in\so(\g)\subset \End(\g)$ and defined by the following properties
\begin{eqnarray}
F(\A,\bt)&\in& \mr{Span}_{\End(\g)}(1,e^{-\A\tau})  \label{F prop 1} , \\
\int_0^1 d\tau\, F(\A,\tau) &=& 0  , \nonumber \\
F(\A,0)&=&1 . \nonumber
\end{eqnarray}
Property (\ref{F prop 1}) is equivalently stated as $(d+\A)F(\A,\bt) = const$ (this constant depends on $\A$). It is convenient to introduce notation
$$R(\A)=-F(\A,1) \in \End(\g) .$$
More explicitly, we have
\begin{eqnarray}
F(\A,\tau)&=&\frac{\frac{1}{2}\left(\coth\frac{\A}{2}+1\right)e^{-\A\tau}-\A^{-1}}{\frac{1}{2}\left(\coth\frac{\A}{2}+1\right)-\A^{-1}} \, ,    \label{F(A)}
\\
R(\A) &=& -\frac{\A^{-1}+\frac{1}{2}-\frac{1}{2}\coth\frac{\A}{2}}{\A^{-1}-\frac{1}{2}-\frac{1}{2}\coth\frac{\A}{2}} \, .
\label{R(A)}
\end{eqnarray}
We will use notation
\be \mu_k(A')=R(\ad_{A_{k-1}}) R(\ad_{A_{k-2}})\cdots R(\ad_{A_{k+1}}) R(\ad_{A_{k}})\in \End(\g)  . \ee
The following reflection properties
$$F(-\A,\tau)=-\frac{F(\A,1-\tau)}{R(\A)},\quad R(-\A)=R(\A)^{-1},\quad \mu_k(A')^T=\mu_k(A')^{-1}$$
mean that $R(\A)$ and $\mu_k(A')$ take values in orthogonal matrices $O(\g)\subset\End(\g)$.

We can now present the result for the integral kernel of the dressed chain homotopy $\kappa_{A'}$:
\begin{lemma}
The operator $\kappa_{A'}$ defined by relations (\ref{K_A prop 1}), (\ref{K_A prop 2}), (\ref{K_A prop 3})
acts on a $\g$-valued 1-form $d\tau\cdot \beta\in \g\otimes\Omega^1(\s^1)$ by formula
\be \kappa_{A'}(d\tau\cdot \beta)(k,\tau)=\sum_{k'=1}^n\int_{\I_{k'}}d\tau'\cdot \kappa_{A'}((k,\tau),(k',\tau'))\circ \beta(k',\tau') , 
\ee
where the integral kernel is given by
\begin{multline}\kappa_{A'}((k,\tau),(k',\tau'))=\\
=\left\{\begin{array}{ll}
\left(\theta(\tau-\tau')-\frac{1}{2}+\frac{1}{2}\coth\frac{\ad_{A_k}}{2}\right)e^{(\tau'-\tau)\ad_{A_k}}-(\ad_{A_k})^{-1}+\\
\qquad \qquad +F(\ad_{A_k},\tau)\left(\frac{1}{1+\mu_k(A')}-\frac{1}{1+R(\ad_{A_k})}\right)F(-\ad_{A_k},\tau') &  \mbox{if}\quad k=k', \\ \\
(-1)^{k-k'} F(\ad_{A_k},\tau)R(\ad_{A_{k-1}})\cdots R(\ad_{A_{k'}})\frac{1}{1+\mu_{k'}(A')} F(-\ad_{A_{k'}},\tau') & \mbox{if}\quad k'<k<k'+n
\end{array}\right. \label{K_A explicit formula}
\end{multline}
This integral kernel is anti-symmetric:
$$\kappa_{A'}((k',\tau'),(k,\tau))^T=-\kappa_{A'}((k,\tau),(k',\tau'))$$
\end{lemma}
To obtain (\ref{K_A explicit formula}) we proceed as in the derivation of (\ref{kappa}). The differential equation implied by (\ref{K_A prop 1}) is as follows:
\be \left(\frac{\dd}{\dd\tau}+\ad_{A_k} \right)\kappa_{A'}((k,\tau),(k',\tau'))=\delta_{k,k'} \delta(\tau-\tau')+C_k(k',\tau') .  \label{K_A eqn} \ee
Conditions (\ref{kappa 2}) are still fulfilled. Solving (\ref{K_A eqn}) with these conditions imposed yields formula (\ref{K_A explicit formula}).

\subsubsection{Simplicial action.}
Simplicial Chern-Simons action $S_{\Xi_n}$ for the triangulation $\Xi_n$ of circle is defined by the fiber BV integral
\be e^{\frac{1}{\hbar}S_{\Xi_n}(\psi',A',\hbar)}=\int_\LL e^{\frac{1}{\hbar}S(\psi'+\psi'',A'+A'')} . \label{fiber BV integral}\ee
As in the case of induction to cohomology, one puts $A''|_\LL=0$,  and the integral becomes  Gaussian:
\be e^{\frac{1}{\hbar}S_{\Xi_n}(\psi',A',\hbar)}=\int\DD\psi''\; e^{\frac{1}{2\hbar}\int_{\s^1}(\psi'+\psi'',d_{A'}(\psi'+\psi''))} . \label{S_Xi functional integral}\ee
Expanding the integrand as
$$(\psi',d_{A'}\psi')+(\psi'',d\psi'')+(\psi'',\ad_{A'}\psi'')+2(\psi'',\ad_{A'}\psi')$$
one can consider the second term as the ``free'' part of the action and the third and fourth terms as a perturbation. In this way, we arrive at the following Feynman diagram expansion for $S_{\Xi_n}$:
\begin{multline}
S_{\Xi_n}=\underbrace{\frac{1}{2}(\psi',d\psi')+\frac{1}{2}(\psi',\ad_{A'}\psi')- \frac{1}{2}(\psi',\ad_{A'}\kappa\,\ad_{A'}\psi')+\frac{1}{2}(\psi',\ad_{A'}\kappa\,\ad_{A'}\kappa\,\ad_{A'}\psi')- \cdots}_{S_{\Xi_n}^0}+\\
+\underbrace{\hbar\frac{1}{2\cdot 2}\,\tr_{\g\otimes\Omega^0(\s^1)}\left(\kappa\,\ad_{A'}\kappa\,\ad_{A'}\right)-\hbar\frac{1}{2\cdot 3}\,\tr_{\g\otimes\Omega^0(\s^1)}\left(\kappa\,\ad_{A'}\kappa\,\ad_{A'}\kappa\,\ad_{A'}\right)+\cdots}_{\hbar S_{\Xi_n}^1}
\label{S_Xi perturbative expansion}
\end{multline}
Here the first line is the sum of ``tree'' diagrams and the second line is the sum of ``wheel diagrams''.

The tree part of $S_{\Xi_n}$ can be  expressed in terms of the dressed chain homotopy,
\be S_{\Xi_n}^0=\frac{1}{2}\int_{\s^1}(\psi',d\psi')+\frac{1}{2}\int_{\s^1}(\psi',\ad_{A'}\psi')- \frac{1}{2}\int_{\s^1}(\psi',\ad_{A'}\kappa_{A'}\ad_{A'}\psi') . \label{S_Xi^0 via K_A}\ee
For the 1-loop part,  we first write
$$S_{\Xi_n}^1=\frac{1}{2}\,\tr_{\g\otimes\Omega^0(\s^1)}\log(1+\kappa\,\ad_{A'}) . $$
Then, by using the general formula
$$\frac{\dd}{\dd s}\tr\, \log M_s= \tr\, \left(M_s^{-1}\frac{\dd M_s}{\dd s}\right)$$
one obtains
$$S_{\Xi_n}^1=\int_0^1 ds\,\frac{\dd}{\dd s}\left(\frac{1}{2}\,\tr_{\g\otimes\Omega^0(\s^1)}\log(1+\kappa\,\ad_{sA'})\right)=
\frac{1}{2}\int_0^1 ds\; \tr_{\g\otimes\Omega^0(\s^1)} \underbrace{(1+\kappa\,\ad_{sA'})^{-1}\kappa}_{\kappa_{sA'}}\ad_{A'} . $$
One can evaluate the functional trace in the integrand in the coordinate representation ({\em i.e.} in the basis of delta-functions) by expressing it in terms of the integral kernel (\ref{K_A explicit formula}) restricted to the diagonal $\s^1\subset \s^1\times \s^1$:
\be S_{\Xi_n}^1=\int_0^1 ds\;\frac{1}{2}\sum_{k=1}^n \int_0^1 d\tau\,\tr_\g \left(\kappa_{sA'}((k,\tau),(k,\tau))\ad_{A_k}\right) . \label{S_Xi^1 via K_A}\ee
There is an ambiguity in this expression since the integral kernel (\ref{K_A explicit formula}) is discontinuous on the diagonal. We regularize this ambiguity by using 
the convention
\be \theta(0)=\frac{1}{2} \label{theta(0)}\ee
Observe that  the value assigned to $\theta(0)$ does not really matter: changing the convention for $\theta(0)$ changes $S_{\Xi_n}^1$ by $\propto\tr_\g \ad_{A'}=0$. It is interesting that the integral over the auxiliary parameter $s$ in (\ref{S_Xi^1 via K_A}) can be computed explicitly.
By putting together (\ref{S_Xi^0 via K_A}) and (\ref{S_Xi^1 via K_A}) and substituting (\ref{K_A explicit formula}) we obtain the following explicit result for the simplicial Chern-Simons action on the triangulated circle.
\begin{theorem}
Simplicial Chern-Simons action on the triangulated circle is given by
\begin{multline}
S_{\Xi_n}=\\
=-\frac{1}{2}\sum_{k=1}^n\left((\psi_k,\psi_{k+1})+\frac{1}{3}(\psi_k,\ad_{A_k}\psi_k)+ \frac{1}{3}(\psi_{k+1},\ad_{A_k}\psi_{k+1})+\frac{1}{3}(\psi_k,\ad_{A_k}\psi_{k+1})\right)+\\
+\frac{1}{2}\sum_{k=1}^n (\psi_{k+1}-\psi_k, \left(\frac{1-R(\ad_{A_k})}{2}\left(\frac{1}{1+\mu_k(A')}-\frac{1}{1+R(\ad_{A_k})}\right) \frac{1-R(\ad_{A_k})}{2 R(\ad_{A_k})}+ \right. \\ \left.
+(\ad_{A_k})^{-1}+ \frac{1}{12}\,\ad_{A_k}-\frac{1}{2}\coth\frac{\ad_{A_k}}{2}\right)\circ(\psi_{k+1}-\psi_k))+\\
+\frac{1}{2}\sum_{k'=1}^n\,\sum_{k=k'+1}^{k'+n-1}(-1)^{k-k'}(\psi_{k+1}-\psi_k, \frac{1-R(\ad_{A_k})}{2}R(\ad_{A_{k-1}})\cdots R(\ad_{A_{k'}})\cdot \\
\cdot\frac{1}{1+\mu_{k'}(\A')}\cdot\frac{1-R(\ad_{A_{k'}})}{2 R(\ad_{A_{k'}})} \circ (\psi_{k'+1}-\psi_{k'}))+\\
+\hbar\,\frac{1}{2}\,\tr_\g\log\left((1+\mu_\bt(A'))\prod_{k=1}^n \left(\frac{1}{1+R(\ad_{A_k})}\cdot\frac{\sinh\frac{\ad_{A_k}}{2}}{\frac{\ad_{A_k}}{2}}\right)\right) .
\label{S_Xi explicit}
\end{multline}
\end{theorem}
In the last term (the 1-loop contribution) $\mu_\bt(A')$ stands for $\mu_l(A')$ for arbitrary $l$.  For different $l$,  these matrices differ by conjugation. Hence, the expression $\det_\g (1+\mu_\bt(A'))$ is well defined.

\subsubsection{Remarks.}
\begin{Rem}
In deriving formula (\ref{S_Xi explicit}) we were sloppy about additive constants (we did not pay attention to normalization of the measure in the functional integral (\ref{S_Xi functional integral})). We chose an ad hoc  normalization
$$S_{\Xi_n}(0,0,\hbar)=-\hbar\, \frac{n-1}{2}\,\dim\g\,\log 2$$
which turns out to be consistent with the operator formalism (see section \ref{sec: op formalism}).
\end{Rem}

\begin{Rem}
Setting $n=1$ in (\ref{S_Xi explicit}) yields the effective action on cohomology (\ref{W on coh of circle}).
\end{Rem}

\begin{Rem}
By expanding (\ref{S_Xi explicit}) in a power series with respect to $A'$ we get back the perturbative expansion  (\ref{S_Xi perturbative expansion})),
\begin{multline}
S_{\Xi_n}=\\
=-\frac{1}{2}\sum_{k=1}^n (\psi_k,\psi_{k+1})- \frac{1}{2}\sum_{k=1}^n\left(\frac{1}{3}(\psi_k,\ad_{A_k}\psi_k)+ \frac{1}{3}(\psi_{k+1},\ad_{A_k}\psi_{k+1})+\frac{1}{3}(\psi_k,\ad_{A_k}\psi_{k+1})\right)+\\
+\frac{1}{2}\sum_{k'=1}^n\,\sum_{k=k'+1}^{k'+n-1}\frac{1}{72}(-1)^{k-k'} (\psi_{k+1}-\psi_k,\; \ad_{A_k}\ad_{A_{k'}}(\psi_{k'+1}-\psi_{k'}))-\\
-\hbar\, \frac{n-1}{2}\,\dim\g\,\log 2
+\hbar\frac{1}{2\cdot 2}\left(\sum_{k=1}^n \frac{1}{12} \tr_\g (\ad_{A_k})^2+ \sum_{k'=1}^n\,\sum_{k=k'+1}^{k'+n-1}\frac{1}{36} \tr_\g(\ad_{A_k}\ad_{A_{k'}})\right)+\mc{O}((A')^3) .
\label{S_Xi series in A'}
\end{multline}
\end{Rem}

\begin{Rem}
Na\"ively, at large $n$ the simplicial action $S_{\Xi_n}$ can be viewed as a lattice approximation to the continuum action (\ref{action on circle}). For $\psi$ and $A$  fixed, we have
$$S_{\Xi_n}\left(\{\psi_k=\psi(k/n)\},\quad \{A_k=\int_{k/n}^{(k+1)/n}A\}\right)\longrightarrow S(\psi,A)+\mc{O}(1/n)$$
when $n$ tends to infinity.
The point is that rather than being just an approximation the simplicial theory $(\FF_{\Xi_n},S_{\Xi_n})$ is exactly equivalent to the continuum theory $(\FF,S)$ for any finite $n$.
\end{Rem}

\begin{Rem}
The simplicial theory is constructed
by the fiber BV integral from the continuum theory. Hence, we expect the simplicial action to satisfy the quantum master equation
\be \Delta_{\Xi_n}e^{\frac{1}{\hbar}S_{\Xi_n}}=0 \label{QME for S_Xi}\ee
with BV Laplacian associated to the BV 2-form (\ref{simplicial omega}),
\be \Delta_{\Xi_n}=\sum_{k=1}^n\frac{\dd}{\dd \tilde\psi_k^a}\frac{\dd}{\dd A_k^a} , \label{Delta_Xi_n}\ee
where
\be\tilde\psi_k=\frac{\psi_k+\psi_{k+1}}{2} . \label{psi tilde definition}\ee
However, the BV Laplacian in the continuum theory is ill-defined.
Therefore, the quantum master equation (\ref{QME for S_Xi}) is not automatic,  and it must be checked independently. It is easy to do it in low degrees in $A'$ by using the expansion (\ref{S_Xi series in A'}). We will prove the quantum master equation via the operator approach in section \ref{sec: QME}.
\end{Rem}

\begin{Rem}
Another property we expect from the simplicial theory is its compatibility with simplicial aggregations $\Xi_{n+2}\ra \Xi_n$. We will prove this property in section \ref{sec: aggregations}.
\end{Rem}

\begin{Rem}
The gauge choice (\ref{PL splitting for Omega}) is actually rigid up to diffeomorphisms of intervals $\I_k$. More exactly, if we require compatibility of the splitting $\Omega=\Omega'\oplus\Omega''$ with the de Rham differential and  with the pairing $\int_{\s^1}\bt\wedge\bt$ ({\em i.e.} so that $\Omega'$ and $\Omega''$ be subcomplexes of $\Omega$, and  $\Omega''$ be orthogonal to $\Omega'$), then the splitting is completely determined by the images of basis 1-cochains of $\Xi_n$ in $\Omega^1(\s^1)$. If in addition, we require that the image of each basis 1-cochain $e_k^{(1)}$  be supported exactly on the respective interval $\I_k\subset \s^1$ (a kind of simplicial locality property), we obtain the splitting (\ref{PL splitting for Omega}) up to diffeomorphisms of intervals $\I_k$ taking the representatives of basis 1-cochains $e_k^{(1)}$ to constant forms $\theta_{\I_k} d\tau$. It is important to note that we are implicitly assuming that the splitting for fields is induced by the splitting for real-valued differential forms. If we drop this assumption and allow splittings of $\g\otimes \Omega$ which are not obtained from a splitting of $\Omega$ by tensoring with $\g$, we can introduce other gauges.
\end{Rem}

\begin{Rem}
The embedding of cell cochains of $\Xi_n$ to the space of differential forms in (\ref{PL splitting for Omega}) is the same as in the 1-dimensional  simplicial BF theory \cite{simpBF}: 0-cochains are represented by continuous piecewise-linear functions, and 1-cochains are represented by piecewise-constant 1-forms. However, the projection (\ref{P'}) is very different. In fact,  it is non-local: for 0-forms instead of evaluation at vertices (as in simplicial BF) we have a sum of integrals over all intervals $\I_k$ with certain signs. Anf  for 1-forms, instead of integration over one interval we have an integral over the whole circle with a certain piecewise-linear integral kernel. Thus, in the splitting $\Omega=\Omega'\oplus\Omega''$ the infrared part is like in the simplicial BF theory, but the ultraviolet part is different.
\end{Rem}

\begin{Rem}
The action (\ref{S_Xi explicit}) can be viewed as a generating function of a certain infinity-structure on $\g\otimes C^\bt(\Xi_n)$. In more detail, this is a structure of a loop-enhanced (or ``quantum'', or ``unimodular'') cyclic $L_\infty$ algebra with the structure maps (operations) $c_k^{(l)}: \wedge^k (\g\otimes C^\bt(\Xi_n))\ra\RR$ related to the action (\ref{S_Xi explicit}) by
$$S_{\Xi_n}=\sum_{l=0}^1 \sum_{k=2}^\infty \frac{\hbar^l}{k!}\; c_k^{(l)}(\underbrace{\psi'+A',\cdots,\psi'+A'}_{k})$$
where the superfield $\psi'+A'$ is understood as the parity-shifted identity map $\FF_{\Xi_n}\ra \g\otimes C^\bt(\Xi_n)$ (as in (\ref{superfield as parity-shifted id})). The quantum master equation (\ref{QME for S_Xi}) generates a family of structure equations on operations $c_k^{(l)}$. In particular,  $c_k^{(0)}$ satisfy the structure equations of the usual (nonunimodular) cyclic $L_\infty$ algebra.
This algebraic structure on $\g$-valued cochains of $\Xi_n$ can be viewed as a homotopy transfer of the unimodular cyclic DGLA structure on $\g\otimes \Omega^\bt(\s^1)$. Only the first two cyclic operations,  $c_2^{(0)}:\wedge^2 (\g\otimes C^\bt(\Xi_n))\ra\RR$ and
$c_3^{(0)}:\wedge^3 (\g\otimes C^\bt(\Xi_n))\ra\RR$
are simplicially-local. All the other 
operations are non-local.
We can also use the pairing on $\g\otimes C^\bt(\Xi_n)$ (induced by the pairing $\int_{\s^1}(\bt,\bt)$ on $\g$-valued differential forms) to invert one input in cyclic operations. This gives an oriented (non-cyclic) version of the unimodular $L_\infty$ structure (see e.g. \cite{Granaker}) on $\g\otimes C^\bt(\Xi_n)$ with structure maps $l_k^{(0)}: \wedge^k (\g\otimes C^\bt(\Xi_n))\ra \g\otimes C^\bt(\Xi_n)$ and $l_k^{(1)}:\wedge^k(\g\otimes C^\bt(\Xi_n))\ra \RR$ related to the cyclic operations $c_{k}^{(l)}$ by the following formula
\begin{eqnarray*}
c_{k+1}^{(0)}(\underbrace{\psi'+A',\cdots,\psi'+A'}_{k+1}) & = & \left(\psi'+A', \; l_k^{(0)}(\underbrace{\psi'+A',\cdots,\psi'+A'}_{k})\right),\qquad k\geq 1 \\
c_k^{(1)} & = & l_k^{(1)}, \qquad k\geq 2
\end{eqnarray*}
In this unimodular $L_\infty$ algebra, only the differential $l_1^{(0)}$ is local while all other operations (including the binary bracket $l_2^{(0)}:\wedge^2 (\g\otimes C^\bt(\Xi_n))\ra \g\otimes C^\bt(\Xi_n)$) are non-local --- unlike in the simplicial BF theory \cite{simpBF} where all higher operations are simplicially-local.
\end{Rem}

\begin{Rem}
The dressed chain homotopy $\kappa_{A'}$ can be used to construct the non-linear map $U:\FF_{\Xi_n}\ra \FF$,
$$U:\psi'+A'\mapsto \psi'+A'-\kappa_{A'}\ad_{A'}\psi'. $$
It sends the infrared field $\psi'+A'$ to the conditional extremum of the continuum action $S$ restricted to $\{\psi'+A'\}\oplus\LL$. The tree part of the simplicial action can be expressed in terms of $U$,
\begin{multline*}
S_{\Xi_n}^0=S(U(\psi'+A'))=\\
=\int_{\s^1}\frac{1}{2} (U(\psi'+A'),d\,U(\psi'+A'))+\frac{1}{6}(U(\psi'+A'),[U(\psi'+A'),U(\psi'+A')])
\end{multline*}
In the language of infinity algebras, $U$ is an $L_\infty$ morphism intertwining the DGLA structure on $\g\otimes\Omega^\bt(\s^1)$ and the $L_\infty$ structure on $\g\otimes C^\bt(\Xi_n)$.
\end{Rem}

\begin{Rem}
There are two natural systems of $\g$-valued coordinates on the space of simplicial BV fields $\FF_{\Xi_n}$: $\{\psi_k,A_k\}$ and $\{\tilde\psi_k,A_k\}$ (where variables $\tilde\psi_k$ are defined by (\ref{psi tilde definition})). The first coordinate system is associated to the realization of the space of fields through the cochain complex of a triangulation: $\FF_{\Xi_n}=\Pi\g\otimes C^\bt(\Xi_n)$. The second coordinate system is associated to the realization through 1-chains and 1-cochains: $\FF_{\Xi_n}\cong \g\otimes C_1(\Xi_n)\oplus \Pi\g\otimes C^1(\Xi_n)$ (or instead of 1-chains of $\Xi_n$ one can talk of 0-cochains of the dual cell decomposition $\Xi_n^\vee$: $\FF_{\Xi_n}\cong \Pi\g\otimes C^0(\Xi_n^\vee)\oplus \Pi\g\otimes C^1(\Xi_n)$). The convenience of the first coordinate system is that the abelian part of the simplcial action (\ref{S_Xi explicit}) is local in variables $(\psi_k,A_k)$. The convenience of the second coordinate system $\{\tilde\psi_k,A_k\}$ is that the BV Laplacian becomes diagonal (\ref{Delta_Xi_n}).
\end{Rem}

%

\section{Approach through operator formalism}
\label{sec 3}

In this section, our strategy is to give a new definition of the one-dimensional Chern-Simons theory. It will be inspired by the definition of section \ref{sec: continuum theory}, but we will be able to consider our theory on an interval and to define a concatenation (gluing) procedure. We will check that the results obtained by the new approach are consistent with those of section \ref{sec: simplicial CS on circle}.

For the rest of the paper, we will assume that $\dim\g=2m$ is even. This is important for theorem \ref{thm: claim} (the correspondence between the operator and path integral formalism): its proof relies on recovering the path integral by using the fundamental representation of Clifford algebra $Cl(\g)$ (section \ref{sec: polarization of g}) which is simpler for $\dim\g=2m$.

\subsection{One-dimensional Chern-Simons theory in operator formalism}
\label{sec: op formalism}

\subsubsection{First approximation}
We would like to define the one-dimensional Chern-Simons theory (\ref{action on circle}) as a quantum mechanics where components of the quantized odd field $\{\hat\psi^a\}$ are subject to the anti-commutation relations
\be \hat\psi^a\hat\psi^b+\hat\psi^b\hat\psi^a=\hbar\,\delta^{ab} , \label{Cl relations}
\ee
{\em i.e.} $\{\hat\psi^a\}$ are generators of the Clifford algebra $Cl(\g)$. The even 1-form (connection) field $A=d\tau\, T^a A^a(\tau)$ is non-dynamical,  and it is treated as a classical background. The evolution operator for the theory on the interval is defined as a path-ordered exponential of $A$ in the spin representation:
\be U_{\I}(A)=\ola{P\exp}\left(-\frac{1}{2\hbar}\int_\I d\tau\, f^{abc}\hat\psi^a A^b(\tau) \hat\psi^c\right)\in Cl(\g) \label{U_I}\ee
Here, the intuition is as follows: the term $\frac{1}{2}\int (\psi, d\psi)$ in the action (\ref{action on circle}) generates the canonical anti-commutation relations (\ref{Cl relations}) while the term $\frac{1}{2}\int (\psi, [A,\psi])$ generates the time-dependent quantum Hamiltonian $\hat{H}(\tau)=-\frac{1}{2} f^{abc}\hat\psi^a A^b(\tau) \hat\psi^c$ which appears in (\ref{U_I}).
The evolution operator for the concatenation of intervals $\I_1=[\p_1,\p_2]$, $\I_2=[\p_2,\p_3]$  with connections $A_1, A_2$ is naturally given by the product of the corresponding evolution operators in the Clifford algebra:
$$U_{[\p_1,\p_3]}(A_1\theta_{[\p_1,\p_2]}+A_2\theta_{[\p_2,\p_3]})=U_{[\p_2,\p_3]}(A_2)\cdot U_{[\p_1,\p_2]}(A_1) .$$
As in section \ref{sec: chain homotopy}, $\theta_{[\p_k,\p_{k+1}]}=\theta_{\I_k}$ denotes the function taking value $1$ on the interval $\I_k$ and zero everywhere else. The partition function for a circle is given by
$$Z_{\s^1}(A)=\Str_{Cl(\g)}\,\ola{P\exp}\left(-\frac{1}{2\hbar}\int_\I d\tau\, f^{abc}\hat\psi^a A^b(\tau) \hat\psi^c\right)\in \RR , $$
where $\Str_{Cl(\g)}$ is the super-trace on $Cl(\g)$ defined as
\be \Str_{Cl(\g)}:\quad \hat{a}\mapsto \left(i\hbar\right)^{m}\cdot\left(\mbox{Coefficient of } \hat\psi^1\cdots\hat\psi^{2m} \;\mbox{ in }\; \hat{a}\right)   . \label{Str_Cl}\ee

\subsubsection{Imposing the cyclic Whitney gauge}
\label{sec: operator formalism, imposing PL gauge}
Our next task is to model in operator formalism the fiber BV integral (\ref{fiber BV integral}) for the theory on circle. We will look for the analogue of the cyclic Whitney gauge introduced in section \ref{sec: PL gauge}. For the connection $A$, imposing the gauge just amounts to saying that $A$ is now infrared, {\em i.e.} a piecewise-constant connection $A'=\sum_{k=1}^n d\tau A_k \theta_{\I_k}$ with $A_k\in\g$. For $\psi$, we would like to restrict the integration to $\psi$'s with given integrals (average values) $\tilde{\psi}_k=\frac{\psi_k+\psi_{k+1}}{2}$ over intervals $\I_k$. So, we are interested in the integral
\begin{multline}
e^{\frac{1}{\hbar}S_{\Xi_n}}=\int\DD\psi\; e^{\frac{1}{2\hbar}\int_{\s^1}\left((\psi,d\psi)+(\psi,\ad_{A'}\psi)\right)} \prod_{k=1}^n\delta\left(\int_{\I_k}d\tau\; \psi-\tilde\psi_k\right)=\\
=\int \prod_{k=1}^n D\lambda_k\; e^{-\sum_{k=1}^n (\lambda_k,\tilde\psi_k)} \underbrace{\int \DD\psi\; e^{\frac{1}{2\hbar}\int_{\s^1}\left((\psi,d\psi)+(\psi,\ad_{A'}\psi)\right)+\sum_{k=1}^n (\lambda_k,\psi)}}_{Z_{\Xi_n}(\lambda',A')}  . \label{BV integral with deltas}
\end{multline}
Here we got rid of $\delta$-functions at the cost of introducing odd auxiliary variables $\lambda_k=T^a\lambda_k^a\in \Pi\g$. One can organize them into an odd piecewise-constant function on the circle $\lambda'=\sum_{k=1}^n \lambda_k \theta_{\I_k}$ which plays the role of  a source for the field $\psi$. The entity $Z_{\Xi_n}(\lambda',A')$ that appeared in the integrand can be written in the operator formalism as
\be Z_{\Xi_n}(\lambda',A')=\Str_{Cl(\g)}\ola{\prod_{k=1}^n}\exp\left(-\frac{1}{2\hbar}f^{abc}\hat\psi^a A_k^b \hat\psi^c+\lambda_k^a \hat\psi^a\right) . \label{Z(lambda,A)}\ee
Then, the partition function of the one-dimensional Chern-Simons theory on the circle (in the Whitney gauge) is given by the odd Fourier transform of (\ref{Z(lambda,A)}):
\be Z_{\Xi_n}(\psi',A')=\left(i\hbar\right)^{-n m}\int \prod_{k=1}^n D\lambda_k\;e^{-\sum_{k=1}^n \lambda_k^a \tilde\psi_k^a}  Z_{\Xi_n}(\lambda',A') . \label{Z(psi,A)}\ee

\begin{Rem} To be precise with signs, we should introduce an ordering convention for the Berezin measure in (\ref{Z(psi,A)}). We set
$$\prod_{k=1}^n D\lambda_k= \ora \prod_{k=1}^n\ora \prod_{a=1}^{2m} D\lambda_k^a .$$
\end{Rem}


\begin{theorem} \label{thm: claim}
For $n$ odd and $\dim\g=2m$ even, one has
\be Z_{\Xi_n}(\psi',A') = e^{\frac{1}{\hbar}S_{\Xi_n}(\psi',A')} , \label{operator - path integral correspondence}\ee
where the right hand side is given by (\ref{S_Xi explicit}) and the left hand side is defined by (\ref{Z(lambda,A)},\ref{Z(psi,A)}).
\end{theorem}

We will prove this theorem in section \ref{sec: back to path integral, non-abelian case} by constructing a  path integral representation for $Z_{\Xi_n}$.
But first (see sections \ref{sec: consistency check 1}, \ref{sec: consistency check 2}) we will perform some direct tests of formula (\ref{operator - path integral correspondence}).

\begin{Rem}
Note that we can define the right hand side of (\ref{operator - path integral correspondence}) only for $n$ odd while the definition of left hand side makes sense for both even and odd$n$. A simple computation shows that for $n$ even the partition function $Z_{\Xi_n}(\psi',A')$ vanishes at  $\psi'=0, A'=0$. This agrees with the observation that for $n$ even the Whitney gauge does not apply (see section \ref{sec: simplicial CS on circle}).
\end{Rem}

\begin{Rem}
We chose the normalization for the super-trace in Clifford algebra (\ref{Str_Cl}) and for $Z_{\Xi_n}$ (\ref{Z(psi,A)})
in a way consistent with the path integral formalism (see (\ref{Phi on product}), (\ref{Str_Cl via eta})).
\end{Rem}

\subsubsection{Consistency check: the effective action on cohomology}\footnote{For reader's convenience, we present explicit calculations in the Clifford algebra here and in subsequent sections; the general reference for the Clifford calculus is \cite{BGV}.}
\label{sec: consistency check 1}
The first test of the correspondence (\ref{operator - path integral correspondence}) is the case of $n=1$. Let us first compute the following expression in the Clifford algebra with two generators $Cl_2$:
\be \varphi(\tilde\psi^1,\tilde\psi^2,a)=(i\hbar)^{-1}\int D\lambda^1 D\lambda^2 \;e^{-\lambda^1 \tilde\psi^1-\lambda^2 \tilde\psi^2} \Str_{Cl_2}\exp\left(-\frac{1}{\hbar}\hat\psi^1 a \hat\psi^2 +\lambda^1 \hat\psi^1+\lambda^2 \hat\psi^2\right) , \label{Cl computation 1}\ee
where  $a\in\RR$ is a number.
For the exponential under the super-trace we have
\begin{multline}
\exp\left(-\frac{1}{\hbar}\hat\psi^1 a \hat\psi^2 +\lambda^1 \hat\psi^1+\lambda^2 \hat\psi^2\right)
=\left(-\frac{2}{\hbar}\sin(a/2)-\frac{\sin (a/2)}{a/2} \lambda^1\lambda^2\right)\hat\psi^1\hat\psi^2+ \\
+\frac{\sin(a/2)}{a/2}(\lambda^1\hat\psi^1+\lambda^2\hat\psi^2)+ \left(\cos(a/2)+\hbar\frac{1}{a}\left(\cos(a/2)-\frac{\sin(a/2)}{a/2}\right)\lambda^1\lambda^2\right) .
\label{Cl exp}
\end{multline}
In this expression, only the first term  contributes to the super-trace in (\ref{Cl computation 1}), and we obtain
\begin{multline*}
\varphi(\tilde\psi^1,\tilde\psi^2,a)= \int D\lambda^1 D\lambda^2 \;e^{-\lambda^1 \tilde\psi^1-\lambda^2 \tilde\psi^2} \left(-\frac{2}{\hbar}\sin(a/2)-\frac{\sin (a/2)}{a/2} \lambda^1\lambda^2\right)=\\
=\frac{\sin (a/2)}{a/2}-\frac{2}{\hbar}\sin(a/2)\tilde\psi^1\tilde\psi^2= \frac{\sin (a/2)}{a/2} e^{-\frac{1}{\hbar}\tilde\psi^1 a\tilde\psi^2} .
\end{multline*}

Next let us consider an anti-symmetric block-diagonal matrix with blocks $2\times 2$
\be \A=\begin{pmatrix}
0 & a^1 & & & & & \\
-a^1 & 0 & & & & & \\
& & 0 & a^2 & & & \\
& & -a^2 & 0 & & & \\
& & & & \ddots & & \\
& & & & & 0 & a^{m} \\
& & & & & -a^{m} & 0
\end{pmatrix}\in \so(\g)\subset \End(\g) .
\label{anti-sym matrix in standard form}
\ee
Then, we have
\begin{multline}
\left(i\hbar\right)^{-m}\int \prod_a D\lambda^a\; e^{-\lambda^a\tilde\psi^a} \Str_{Cl(\g)}\exp\left(-\frac{1}{2\hbar}\hat\psi^a \A^{ab} \hat\psi^b+\lambda^a\hat\psi^a\right)=\\
=\prod_{p=1}^{m}\varphi(\tilde\psi^{2p-1},\tilde\psi^{2p},a^p)= {\det}_g^{1/2}\left(\frac{\sinh\frac{\A}{2}}{\frac{\A}{2}}\right)\cdot e^{-\frac{1}{2\hbar}\tilde\psi^a\A^{ab}\tilde\psi^b} .
\label{Cl computation 2}
\end{multline}
Since both the left hand side and the right hand side of (\ref{Cl computation 2}) are $SO(\g)$-invariant, the equality actually holds for all anti-symmetric matrices $\A$, and in particular for $\A=\ad_{A_0}$, where $A_0\in\g$. Thus, we have shown that
$$Z_{\Xi_1}(\tilde\psi,A_0)=
{\det}_g^{1/2}\left(\frac{\sinh\frac{\ad_{A_0}}{2}}{\frac{\ad_{A_0}}{2}}\right)\cdot e^{-\frac{1}{2\hbar}(\tilde\psi,\ad_{A_0}\tilde\psi)}$$
The right hand side coincides with (\ref{W on coh of circle}), so we have checked the correspondence (\ref{operator - path integral correspondence}) in the case of $n=1$.

\subsubsection{Consistency check: the case of mutually commuting $\{A_k\}$ and $\psi'=0$}
\label{sec: consistency check 2}
Now we would like to perform a direct check of the correspondence (\ref{operator - path integral correspondence}) at the point $\psi'=0$ ({\em i.e.} neglecting the tree part of the simplicial action) and assuming that $[A_k,A_{k'}]=0$ for all $k,k'=1,\ldots,n$. That is, we will check that
\begin{multline}
\left(i\hbar\right)^{-nm} \int \prod_{k=1}^n D\lambda_k \; \Str_{Cl(\g)}\ola{\prod_{k=1}^n}\exp\left(-\frac{1}{2\hbar}f^{abc}\hat\psi^a A_k^b \hat\psi^c+\lambda_k^a \hat\psi^a\right)=\\
=
{\det}_\g^{1/2}\left(\frac{1+\prod_{k=1}^n R(\ad_{A_k})}{\prod_{k=1}^n(1+R(\ad_{A_k}))}\cdot\prod_{k=1}^n\frac{\sinh\frac{\ad_{A_k}}{2}}{\frac{\ad_{A_k}}{2}}\right) .
\label{consistency 1}
\end{multline}
We use the idea of section \ref{sec: consistency check 1} to reduce (\ref{consistency 1}) to a computation in $Cl_2$. Since all $A_k$ mutually commute, we can choose an orthonormal basis in $\g$, such that the matrices $\ad_{A_k}$ simultaneously assume the standard form (\ref{anti-sym matrix in standard form}). Then, both sides of (\ref{consistency 1}) factorize into contributions of $2\times 2$ blocks, and it suffices to check the identity
\begin{multline}
\left(i\hbar\right)^{-n} \int \prod_{k=1}^n (D\lambda_k^1 D\lambda_k^2) \; \Str_{Cl_2}\ola{\prod_{k=1}^n}\exp\left(-\frac{1}{\hbar}\hat\psi^1 a_k\hat\psi^2+\lambda_k^1 \hat\psi^1+\lambda_k^2 \hat\psi^2\right)=\\
=
{\det}^{1/2}\left(\frac{1+\prod_{k=1}^n R(i\sigma_2 a_k)}{\prod_{k=1}^n(1+R(i\sigma_2 a_k))}\cdot\prod_{k=1}^n\frac{\sinh\frac{i\sigma_2 a_k}{2}}{\frac{i\sigma_2 a_k}{2}}\right) .
\label{consistency 2}
\end{multline}
Here $\sigma_2=\begin{pmatrix} 0 & -i \\ i & 0 \end{pmatrix}$ is the second Pauli matrix, and $i\sigma_2 a_k=\begin{pmatrix} 0 & a_k \\ -a_k & 0 \end{pmatrix} $.
To evaluate the left hand side of (\ref{consistency 2}), we use the result (\ref{Cl exp}) for the  Clifford exponential:
$$\mbox{l.h.s of }(\ref{consistency 2})= \left(i\hbar\right)^{-n} \Str_{Cl_2} \ola{\prod_{k=1}^n} \left(\frac{\sin(a_k/2)}{a_k/2}\hat\psi^1\hat\psi^2+\hbar \frac{1}{a_k} \left(\frac{\sin(a_k/2)}{a_k/2}-\cos(a_k/2)\right)\right) $$
 An easy way to evaluate this expression is to use the matrix representation $Cl_2\ra \End(\CC^{1|1})$ which maps
$$\hat\psi^1\mapsto \left(\frac{\hbar}{2}\right)^{1/2}\left(\begin{array}{l|l}0 & 1 \\ \hline 1 & 0 \end{array}\right),\quad \hat\psi^2\mapsto \left(\frac{\hbar}{2}\right)^{1/2}\left(\begin{array}{l|l}0 & -i \\ \hline i & 0 \end{array}\right),\quad \Str_{Cl_2}\mapsto 
\Str_{\End(\CC^{1|1})}, $$
and
$$\Str_{\End(\CC^{1|1})}: \left(\begin{array}{l|l}\alpha & \beta \\ \hline \gamma & \delta \end{array}\right)\mapsto \alpha-\delta$$
is the standard super-trace on matrices of the size $(1|1)\times (1|1)$. Using this representation, we obtain
\begin{multline*}
\mbox{l.h.s of }(\ref{consistency 2})= \\
i^{-n}\Str_{\End(\CC^{1|1})} \prod_{k=1}^n \left(\begin{array}{l|l}\frac{1}{a_k} \left(\frac{\sin(a_k/2)}{a_k/2}-\cos(a_k/2)\right)+\frac{i}{2} \frac{\sin(a_k/2)}{a_k/2} & 0 \\ \hline 0 & \frac{1}{a_k} \left(\frac{\sin(a_k/2)}{a_k/2}-\cos(a_k/2)\right)-\frac{i}{2} \frac{\sin(a_k/2)}{a_k/2} \end{array}\right)\\
=i^{-n}\prod_{k=1}^n \frac{\sin(a_k/2)}{a_k/2} \cdot 2i \, \mr{Im}\prod_{k=1}^n\left(\frac{i}{2}+\frac{1}{a_k}-\frac{1}{2}\cot(a_k/2)\right)=\\
=2 \prod_{k=1}^n \frac{\sin(a_k/2)}{a_k/2} \cdot \mr{Re} \prod_{k=1}^n \frac{1}{1+R(i a_k)} .
\end{multline*}
In the last line we used the assumption that $n$ is odd.

To evaluate the right hand side of (\ref{consistency 2}), first observe that the matrix
$$\frac{1+\prod_{k=1}^n R(i\sigma_2 a_k)}{\prod_{k=1}^n(1+R(i\sigma_2 a_k))}= \frac{1}{\prod_{k=1}^n(1+R(i\sigma_2 a_k))}+\frac{1}{\prod_{k=1}^n(1+R(-i\sigma_2 a_k))}$$
is constructed from matrices $i\sigma_2 a_k=\begin{pmatrix} 0 & a_k \\ -a_k & 0 \end{pmatrix}$ and is therefore of form $\begin{pmatrix} \alpha & \beta \\ -\beta & \alpha \end{pmatrix}$ ({\em i.e.} belongs to $\mr{Span}_\RR(1,i\sigma_2)$), and that it is symmetric. Hence, it is actually a multiple of the identity matrix $\begin{pmatrix} \alpha & 0 \\ 0 & \alpha \end{pmatrix}$, and the determinant may be expressed  as
\begin{multline*}
{\det}^{1/2}\left(\frac{1+\prod_{k=1}^n R(i\sigma_2 a_k)}{\prod_{k=1}^n(1+R(i\sigma_2 a_k))}\right) = \frac{1}{2}\;\tr\left(\frac{1+\prod_{k=1}^n R(i\sigma_2 a_k)}{\prod_{k=1}^n(1+R(i\sigma_2 a_k))}\right)=\\ =
\prod_{k=1}^n \frac{1}{1+R(i a_k)}+\prod_{k=1}^n \frac{1}{1+R(-i a_k)} = 2\,\mr{Re} \prod_{k=1}^n \frac{1}{1+R(i a_k)}
\end{multline*}
Now it is obvious that (\ref{consistency 2}) is verified, and this implies (\ref{consistency 1}).


\subsection{One-dimensional Chern-Simons with boundary}

\subsubsection{One-dimensional simplicial Chern-Simons in the operator formalism. Concatenations.}
\label{sec: operator formalism, 1d simp cs}
There is a natural construction of the one-dimensional simplicial Chern-Simons theory in the operator formalism: to a one-dimensional oriented simplicial complex $\Theta$ (a collection of $i(\Theta)$ triangulated intervals and $c(\Theta)$ circles) it associates a partition function
\be Z_\Theta\in \Fun(\underbrace{\g\otimes C_1(\Theta)\oplus \Pi\g\otimes C^1(\Theta)}_{\FF^\bulk_\Theta}) \otimes Cl(\g)^{\otimes i(\Theta)} , \label{Z_Theta}\ee
where we think of variables $A_k\in \g$ as coordinates on $\g$-valued simplicial 1-cochains of $\Theta$ with parity shifted, and we think of variables $\tilde\psi_k\in \Pi\g$ as coordinates on $\g$-valued simplicial 1-chains of $\Theta$. The partition function $Z_\Theta$ is defined by the following properties:
\begin{itemize}
\item For a single interval with a standard triangulation, we have
\be Z_\I= \left(i\hbar\right)^{-m}\int\; \prod_{a=1}^{2m} D\lambda^a\;\; e^{-\lambda^a\tilde\psi^a}\exp\left(-\frac{1}{2\hbar}f^{abc}\hat\psi^a A^b \hat\psi^c+\lambda^a \hat\psi^a\right) . \label{Z_I}\ee
\item For a disjoint union of $\Theta_1$ and $\Theta_2$,
$$ Z_{\Theta_1\sqcup\, \Theta_2}= Z_{\Theta_1}\otimes Z_{\Theta_2} . $$
\item For a concatenation $\Theta_1\cup \Theta_2$ with intersection $\Theta_1\cap \Theta_2=\p$ being a single point embedded as a boundary point of positive orientation $\p_1\in \Theta_1$ belonging to the $i$-th triangulated interval of $\Theta_1$, and embedded as a boundary point of negative orientation $\p_2\in \Theta_2$ belonging to the $j$-th triangulated interval of $\Theta_2$, we have:
\be Z_{\Theta_1\cup \Theta_2} = \m(Z_{\Theta_2}\otimes Z_{\Theta_1}) . \label{Z of concatenation}\ee
Here $\m: Cl(\g)\otimes Cl(\g)\ra Cl(\g)$ is the Clifford algebra multiplication, and $\m$ in (\ref{Z of concatenation})  acts on Clifford algebras associated to the $j$-th triangulated interval of $\Theta_2$ and the $i$-th triangulated interval of $\Theta_1$.
\item If the simplicial complex $\Theta'$ is obtained from $\Theta$ by closing the $i$-th triangulated interval into a circle, we have:
\be Z_{\Theta'}=\Str_{Cl(\g)} Z_{\Theta} , \label{Z of closure}\ee
where the super-trace is taken over the Clifford algebra associated to the $i$-th interval of $\Theta$.
\end{itemize}

Obviously, this construction gives (\ref{Z(psi,A)}) for a triangulated circle $\Theta=\Xi_n$,  and due to the correspondence (\ref{operator - path integral correspondence}) it is consistent with the results of section \ref{sec: simplicial CS on circle}.

\begin{Rem}
We may regard $Z_\I$ as a contribution of an open interval. Then, the concatenation (\ref{Z of concatenation}) is understood as taking a disjoint union of open triangulated intervals and then gluing them together at the point $\p$. Likewise, (\ref{Z of closure}) is understood as gluing together of two end-points of the same interval. Thus, a contribution of point is either the (1,2)-tensor\footnote{We mean the rank of the tensor: one time covariant, two times contravariant.} $\m$ on the Clifford algebra or the (0,1)-tensor $\Str_{Cl(\g)}$, depending on whether we are gluing together two different or the same connected component.
\end{Rem}

\begin{Rem} Once we come to a description of the one-dimensional Chern-Simons theory in terms of Atiyah-Segal's axioms and specify the vector space associated to a point (the space of states), properties (\ref{Z of concatenation}) and (\ref{Z of closure})  become a single sewing axiom (see section \ref{sec: Segal's axioms} below).
\end{Rem}

\subsubsection{Simplicial aggregations}
\label{sec: aggregations}
Let $\Theta'$ be the interval $[\p_1,\p_3]$ with the standard triangulation, and let $\Theta$ be the subdivision of $\Theta'$ with three 0-simplices $\p_1,\p_2,\p_3$ and two 1-simplices $[\p_1,\p_2],\;[\p_2,\p_3]$. The aggregation morphism $r^\xi$ acts on simplicial chains as
\begin{multline*}
r^\xi_{C_\bt}:\;C_\bt(\Theta)\ra C_\bt(\Theta') \\
\alpha_{\p_1}e^{\p_1}+\alpha_{\p_2}e^{\p_2}+\alpha_{\p_3}e^{\p_3}+ \alpha_{\p_1\p_2}e^{\p_1\p_2}+\alpha_{\p_2\p_3}e^{\p_2\p_3}\mapsto \\
\mapsto(\alpha_{\p_1}+(1-\xi)\alpha_{\p_2})e'^{\p_1}+ (\alpha_{\p_3}+\xi\alpha_{\p_2})e'^{\p_3}+(\xi \alpha_{\p_1\p_2}+(1-\xi)\alpha_{\p_2\p_3})e'^{\p_1\p_3} ,
\end{multline*}
and on simplicial cochains as
\begin{multline*}
r^\xi_{C^\bt}:\;C^\bt(\Theta)\ra C^\bt(\Theta') \\
\alpha^{\p_1}e_{\p_1}+\alpha^{\p_2}e_{\p_2}+\alpha^{\p_3}e_{\p_3}+ \alpha^{\p_1\p_2}e_{\p_1\p_2}+\alpha^{\p_2\p_3}e_{\p_2\p_3}
\mapsto \alpha^{\p_1} e'_{\p_1}+\alpha^{\p_3}e'_{\p_3}+ (\alpha^{\p_1\p_2}+\alpha^{\p_2\p_3})e'_{\p_1\p_3} .
\end{multline*}
Here $0<\xi<1$ is a parameter determining the relative weight of intervals $[\p_1,\p_2]$ and $[\p_2,\p_3]$ inside $[\p_1,\p_3]$ (the symmetric choice corresponds to $\xi=1/2$). Basis chains and cochains corresponding a simplex $\sigma$ are denoted by $e^\sigma$ and $e_\sigma$, respectively. We use primes to distinguish the basis of
$C_\bt(\Theta'), C^\bt(\Theta')$ ;
${\alpha_\mr{\ldots}}, {\alpha^\mr{\ldots}}$ are numerical coefficients.

One can also introduce the subdivision morphism which is dual to the aggregation morphism:
$$i^\xi_{C_\bt}=(r^\xi_{C^\bt})^*:\; C_\bt(\Theta')\ra C_\bt(\Theta),\quad i^\xi_{C^\bt}=(r^\xi_{C_\bt})^*:\; C^\bt(\Theta')\ra C^\bt(\Theta) .$$
More explicitly,
\begin{multline*}
i^\xi_{C_\bt}:\; \alpha'_{\p_1}e'^{\p_1}+\alpha'_{\p_3}e'^{\p_3}+\alpha'_{\p_1\p_3}e'^{\p_1\p_3}\mapsto \alpha'_{\p_1}e^{\p_1}+ \alpha'_{\p_3}e^{\p_3}+ \alpha'_{\p_1\p_3} (e^{\p_1\p_2}+e^{\p_2\p_3}), \\
i^\xi_{C^\bt}:\; \alpha'^{\p_1}e'_{\p_1}+\alpha'^{\p_3}e'_{\p_3}+\alpha'^{\p_1\p_3}e'_{\p_1\p_3}\mapsto \\
\mapsto\alpha'^{\p_1} e_{\p_1}+((1-\xi)\alpha'^{\p_1}+\xi\alpha'^{\p_3}) e_{\p_2}+\alpha'^{\p_3} e_{\p_3}+ \alpha'^{\p_1\p_3} (\xi e_{\p_1\p_2}+ (1-\xi) e_{\p_2\p_3}) .
\end{multline*}
Aggregation and subdivision morphisms are chain maps. Moreover, they are quasi-isomorphisms.

These maps induce an embedding
$i^\xi_{\Theta,\Theta'}: \FF_{\Theta'}^\bulk\hra \FF_{\Theta}^\bulk$ and projection $r^\xi_{\Theta,\Theta'}: \FF_{\Theta}^\bulk\ra \FF_{\Theta'}^\bulk$ for the spaces of ``bulk fields''
$$\FF_{\Theta}^\bulk=\g\otimes C_1(\Theta)\oplus \Pi\g\otimes C^1(\Theta),\quad \FF_{\Theta'}^\bulk=\g\otimes C_1(\Theta')\oplus \Pi\g\otimes C^1(\Theta') .$$
That is,  we have a splitting
\be \FF_{\Theta}^\bulk=i^\xi_{\Theta,\Theta'}(\FF_{\Theta'}^\bulk)\oplus \FF''^{\bulk,\xi}_{\Theta,\Theta'} .  \label{aggregation - splitting}\ee
In more detail, we split the bulk fields as
$$\tilde\psi_1=\tilde\psi'+(1-\xi)\tilde\psi'',\quad \tilde\psi_2=\tilde\psi'-\xi\tilde\psi'', \quad A_1=\xi A'-A'' ,\quad A_2=(1-\xi)A'+A''.$$
The maps $i^\xi_{\Theta,\Theta'}$ and $r^\xi_{\Theta,\Theta'}$ are dual to each other with respect to the odd pairing on $\FF_{\Theta}^\bulk$ and $\FF_{\Theta'}^\bulk$ (thus, (\ref{aggregation - splitting}) is an orthogonal decomposition).
We denote by $\LL^\xi_{\Theta,\Theta'}\subset \FF''^{\bulk,\xi}_{\Theta,\Theta'}$ the Lagrangian subspace defined by setting to zero the 1-cochain part of the ultraviolet field $A''$.

We define the action of the aggregation map $\Theta\ra \Theta'$ on the partition function of the one-dimensional simplicial Chern-Simons theory by the fiber BV integral associated to the splitting (\ref{aggregation - splitting}):
\be (r_{\Theta,\Theta'}^\xi)_*(Z_\Theta)=\int_{\LL^\xi_{\Theta,\Theta'}}Z_\Theta \quad \in \Fun(\FF_{\Theta'}^\bulk)\otimes Cl(\g) . \label{aggregation - BV integral}\ee

It is easy to give a direct check of the following statement.
\begin{lemma}
\be (r_{\Theta,\Theta'}^\xi)_* (Z_\Theta) = Z_{\Theta'} \label{aggregation - compatibility}\ee
\end{lemma}

\textit{Proof.} Indeed, by definition (\ref{aggregation - BV integral}) we have
\begin{multline*}(r_{\Theta,\Theta'}^\xi)_* (Z_\Theta)=\\
=\left(i\hbar\right)^{m} \int \ola{\prod_{a=1}^{2m}}D\tilde\psi''^a\;Z_{[\p_2,\p_3]}\left(\tilde\psi'-\xi\tilde\psi'',(1-\xi)A'\right)\cdot Z_{[\p_1,\p_2]}\left(\tilde\psi'+(1-\xi)\tilde\psi'',\xi A'\right)= \\
=\left(i\hbar\right)^{-m}\int \ola{\prod_{a=1}^{2m}}D\tilde\psi''^a\, \ora{\prod_{a=1}^{2m}} D\lambda_2^a \,\ora{\prod_{a=1}^{2m}}  D\lambda_1^a\; e^{-\lambda_1^a (\tilde\psi'^a+(1-\xi)\tilde\psi''^a)-\lambda_2^a (\tilde\psi'^a-\xi\tilde\psi''^a)}\cdot \\
\cdot\exp\left(-\frac{1-\xi}{2\hbar}f^{abc}\hat\psi^a A'^b\hat\psi^c+\lambda_2^a \hat\psi^a\right)\cdot \exp\left(-\frac{\xi}{2\hbar}f^{abc}\hat\psi^a A'^b\hat\psi^c+\lambda_1^a \hat\psi^a\right) .
\end{multline*}
Here we made a change of coordinates $(\lambda_1,\lambda_2) \ra (\lambda=\lambda_1+\lambda_2, \nu=(1-\xi)\lambda_1-\xi\lambda_2)$. The integral over $\tilde\psi''$ produces the delta-function $\delta(\nu)$; by integrating over $\nu$ we obtain
\begin{multline*}
(r_{\Theta,\Theta'}^\xi)_* (Z_\Theta)
 =\left(i\hbar\right)^{-m} \int \ora{\prod_{a=1}^{2m}}  D\lambda^a\; e^{-\lambda^a \tilde\psi'^a}\cdot\\
 \cdot\exp\left(-\frac{1-\xi}{2\hbar}f^{abc}\hat\psi^a A'^b\hat\psi^c+(1-\xi)\lambda^a \hat\psi^a\right)\cdot \exp\left(-\frac{\xi}{2\hbar}f^{abc}\hat\psi^a A'^b\hat\psi^c+\xi \lambda^a \hat\psi^a\right)=\\
 =\left(i\hbar\right)^{-m} \int \ora{\prod_{a=1}^{2m}}  D\lambda^a\; e^{-\lambda^a \tilde\psi'^a}
 \cdot\exp\left(-\frac{1}{2\hbar}f^{abc}\hat\psi^a A'^b\hat\psi^c+\lambda^a \hat\psi^a\right)=Z_{\Theta'} .
\end{multline*}
$\Box$

\begin{Rem} We normalize the measure on $\LL^\xi_{\Theta,\Theta'}$ in such a way that relation (\ref{aggregation - compatibility}) holds with no additional factors.
\end{Rem}

Up to now, we only discussed the elementary aggregation which takes an interval subdivided into two smaller intervals and into an interval with the standard triangulation (that is, one removes the middle point $\p_2$ and merges intervals $[\p_1,\p_2]$ and $[\p_2,\p_3]$). A general simplicial aggregation for a one-dimensional simplicial complexe is a sequence of elementary aggregations made at each step on an incident pair of intervals. In particular, there are many simplicial aggregations for triangulated circles: $\Xi_n\ra \Xi_{n'}$ with $n> n'$.

The following is an immediate consequence of (\ref{aggregation - compatibility}):
\begin{proposition}
For a general simplicial aggregation
$$r=r_{\Theta_{l-1},\Theta'}^{\xi_l}\circ\cdots \circ r_{\Theta_1,\Theta_2}^{\xi_2}\circ r_{\Theta,\Theta_1}^{\xi_1}:\;\Theta\ra\Theta'$$
(where $\Theta$ is an arbitrary one-dimensional simplicial complex and $\Theta'$ is some aggregation of $\Theta$), one has
\be r_*(Z_\Theta)=Z_{\Theta'} . \ee
\end{proposition}
The compatibility with aggregations 
is an important property expected from a simplicial theory. In particular, (\ref{operator - path integral correspondence}) implies that the simplicial action for a circle $S_{\Xi_n}$ given by (\ref{S_Xi explicit}) is compatible with simplicial aggregations $\Xi_n\ra \Xi_{n'}$ for $n,n'$ odd.

\subsubsection{Quantum master equation}
\label{sec: QME}
\begin{lemma} \label{lemma: QME}\footnote{In a different setting equation (\ref{QME for Z_I}) appeared in \cite{Alekseev-Meinrenken}.}
The partition function for an interval (\ref{Z_I}) satisfies the following differential equation:
\be \hbar\, \frac{\dd}{\dd\tilde\psi^a}\,\frac{\dd}{\dd A^a}Z_\I+ \frac{1}{\hbar}\, \left[\frac{1}{6} f^{abc} \hat\psi^a\hat\psi^b\hat\psi^c,Z_\I\right]_{Cl(\g)}=0 , \label{QME for Z_I}\ee
where $[,]_{Cl(\g)}$ denotes the super-commutator on $Cl(\g)$.
\end{lemma}

\textit{Proof.} We will check (\ref{QME for Z_I}) using variables $(\lambda,A)$. We have,
\be \left(\hbar\, \lambda^a\frac{\dd}{\dd A^a}+  \frac{1}{\hbar}\, \left[\frac{1}{6} f^{abc} \hat\psi^a\hat\psi^b\hat\psi^c,\bt\right]_{Cl(\g)}\right) \exp\left(-\frac{1}{2\hbar}f^{abc}\hat \psi^a A^b \hat\psi^c+\lambda^a \hat\psi^a\right)=0 .\label{QME eq1} \ee
After the Fourier transform from the variable $\lambda$ to the variable $\tilde\psi$, this expression becomes (\ref{QME for Z_I})). Observe that
\begin{multline}
\hbar\, \lambda^a\frac{\dd}{\dd A^a} \exp\left(-\frac{1}{2\hbar}f^{abc}\hat \psi^a A^b \hat\psi^c+\lambda^a \hat\psi^a\right)= \\
=\int_0^1 d\tau\; e^{\tau \left(-\frac{1}{2\hbar }f^{abc}\hat \psi^a A^b \hat\psi^c+\lambda^a \hat\psi^a\right)}\cdot \frac{1}{2}f^{abc}\hat \psi^a \lambda^b \hat\psi^c\cdot e^{(1-\tau) \left(-\frac{1}{2\hbar}f^{abc}\hat \psi^a A^b \hat\psi^c+\lambda^a \hat\psi^a\right)} .
\label{QME eq2}
\end{multline}
Next, compute commutators in the Clifford algebra,
\be \left[\frac{1}{6} f^{abc} \hat\psi^a\hat\psi^b\hat\psi^c,\lambda^{a'} \hat\psi^{a'}\right]_{Cl(\g)}=-\frac{\hbar}{2}f^{abc}\hat \psi^a \lambda^b \hat\psi^c , \label{Cl commutator 1}\ee
and
\begin{multline}\left[\frac{1}{6} f^{abc} \hat\psi^a\hat\psi^b\hat\psi^c,\frac{1}{2}f^{a'b'c'}\hat \psi^{a'} A^{b'} \hat\psi^{c'}\right]_{Cl(\g)}=\\
=\frac{1}{12}f^{abc}f^{a'b'c'} A^{b'}([\hat\psi^a\hat\psi^b\hat\psi^c,\hat\psi^{a'}]\hat\psi^{c'}- \hat\psi^{a'}[\hat\psi^a\hat\psi^b\hat\psi^c,\hat\psi^{c'}])=\\
=\frac{\hbar}{4} f^{abc}f^{a'b'c'} A^{b'} (\delta^{c a'}\hat\psi^a\hat\psi^b\hat\psi^{c'}-\delta^{cc'}\hat\psi^{a'}\hat\psi^a\hat\psi^b)=
\frac{\hbar}{4} f^{abc}f^{c b'c'} A^{b'} (\hat\psi^a\hat\psi^b\hat\psi^{c'}+\hat\psi^{c'}\hat\psi^a\hat\psi^b)=\\
=\underbrace{\frac{\hbar}{6} f^{abc}f^{c b'c'} A^{b'} (\hat\psi^a\hat\psi^b\hat\psi^{c'}+\hat\psi^{c'}\hat\psi^a\hat\psi^b+\hat\psi^b\hat\psi^{c'}\hat\psi^a)}_{=0\mbox{ by Jacobi identity}}+\\
+\frac{\hbar}{12} f^{abc}f^{c b'c'} A^{b'} (\hat\psi^a\hat\psi^b\hat\psi^{c'}+\hat\psi^{c'}\hat\psi^a\hat\psi^b- 2\hat\psi^b\hat\psi^{c'}\hat\psi^a)=\\
=\frac{\hbar}{24} f^{abc}f^{c b'c'} A^{b'} ((\hat\psi^a\hat\psi^b\hat\psi^{c'}+\hat\psi^{c'}\hat\psi^a\hat\psi^b- 2\hat\psi^b\hat\psi^{c'}\hat\psi^a)-(\hat\psi^b\hat\psi^a\hat\psi^{c'}+\hat\psi^{c'}\hat\psi^b\hat\psi^a- 2\hat\psi^a\hat\psi^{c'}\hat\psi^b))=\\
=\frac{\hbar}{24} f^{abc}f^{c b'c'} A^{b'} (\hat\psi^a [\hat\psi^b,\hat\psi^{c'}]+[\hat\psi^{c'},\hat\psi^a]\hat\psi^b-\hat\psi^b[\hat\psi^{c'},\hat\psi^a]-[\hat\psi^b,\hat\psi^{c'}] \hat\psi^a)=0 .
\label{Cl commutator 2}
\end{multline}
For brevity, we are omitting the subscript in $[,]_{Cl(\g)}$ in computations. Identities (\ref{Cl commutator 1},\ref{Cl commutator 2}) imply
\begin{multline*}
 \frac{1}{\hbar}\, \left[\frac{1}{6} f^{abc} \hat\psi^a\hat\psi^b\hat\psi^c,\exp\left(-\frac{1}{2\hbar}f^{abc}\hat \psi^a A^b \hat\psi^c+\lambda^a \hat\psi^a\right)\right]_{Cl(\g)}=\\
=-\int_0^1 d\tau\; e^{\tau \left(-\frac{1}{2\hbar}f^{abc}\hat \psi^a A^b \hat\psi^c+\lambda^a \hat\psi^a\right)}\cdot \frac{1}{2}f^{abc}\hat \psi^a \lambda^b \hat\psi^c\cdot e^{(1-\tau) \left(-\frac{1}{2\hbar}f^{abc}\hat \psi^a A^b \hat\psi^c+\lambda^a \hat\psi^a\right)} .
\end{multline*}
Together with (\ref{QME eq2}), this implies (\ref{QME eq1}) which finishes the proof of (\ref{QME for Z_I}).
$\Box$

Let us denote\footnote{The notation stems from the fact that this is a quantization of the Maurer-Cartan element $\theta\in \Fun(\Pi\g)$ (\ref{theta}).}
by $$\hat \theta:=\frac{1}{6} f^{abc} \hat\psi^a\hat\psi^b\hat\psi^c$$ the Clifford element in (\ref{QME for Z_I}).

\begin{proposition}
The partition  function for any one-dimensional simplicial complex $\Theta$ satisfies the differential equation
\be (\hbar\, \Delta_\Theta^\bulk+\frac{1}{\hbar}\, \delta_\Theta)Z_\Theta=0 , \label{QME for Z_Theta}\ee
where
$$\Delta_\Theta^\bulk=\sum_{k}\frac{\dd}{\dd\tilde\psi_k^a}\frac{\dd}{\dd A_k^a}$$
(the sum goes over all 1-simplices of $\Theta$), and
$$\delta_\Theta=\sum_{j=1}^{i(\Theta)} \left[\hat \theta^{(j)},\bt\right]_{Cl(\g)} .$$
Here the sum goes over connected components of $\Theta$  that are triangulated intervals; $\hat \theta^{(j)}$ denotes $\hat \theta$ as an element of the $j$-th copy of $Cl(\g)$.
\end{proposition}

\textit{Proof.} Equation (\ref{QME for Z_Theta}) follows from (\ref{QME for Z_I}). The compatibility with disjoint unions is obvious as $\Delta_{\Theta_1\sqcup\,\Theta_2}^\bulk=\Delta_{\Theta_1}^\bulk+\Delta_{\Theta_2}^\bulk$, and $\delta_{\Theta_1\sqcup\,\Theta_2}=\delta_{\Theta_1}+\delta_{\Theta_2}$. For concatenations, it suffices to check the case of two triangulated intervals $\Theta_1$ and  $\Theta_2$:
\begin{multline*}
(\hbar \Delta_{\Theta_1\cup\Theta_2}^\bulk+\hbar^{-1}\delta_{\Theta_1\cup\Theta_2})Z_{\Theta_1\cup\Theta_2}= \left(\hbar\Delta_{\Theta_1}^\bulk+\hbar\Delta_{\Theta_2}^\bulk+\hbar^{-1}\left[\hat \theta,\bt\right]_{Cl(\g)}\right)\circ (Z_{\Theta_2}\cdot Z_{\Theta_1})=\\ =
Z_{\Theta_2}\cdot \left(\left(\hbar \Delta_{\Theta_1}^\bulk+\hbar^{-1}\left[\hat \theta,\bt\right]_{Cl(\g)}\right)\circ Z_{\Theta_1}\right)
+
\left(\left(\hbar\Delta_{\Theta_2}^\bulk +\hbar^{-1}\left[\hat \theta,\bt\right]_{Cl(\g)}\right)\circ Z_{\Theta_2}\right)\cdot Z_{\Theta_1}=0 .
\end{multline*}
The compatibility with closure of a triangulated interval into a triangulated circle follows from $\Str_{Cl(\g)}[\hat \theta,Z_\Theta]_{Cl(\g)}=0$.
$\Box$

\begin{Rem} We understand (\ref{QME for Z_Theta}) as a kind of quantum master equation with the boundary term $\delta_\Theta Z_\Theta$.  It is tempting to think of the operator $\Delta_\Theta^\bulk+\delta_\Theta$ appearing in (\ref{QME for Z_Theta}) as a new BV Laplacian adjusted for the presence of the boundary.
\end{Rem}

\begin{corollary} In the case of a triangulated circle $\Theta=\Xi_n$, the partition function satisfies the usual (non modified) quantum master equation
$$\Delta_{\Xi_n}Z_{\Xi_n}=0.$$
\end{corollary}

\section{Back to path integral}

Sections \ref{sec: polarization of g} and \ref{sec: back to PI} mostly go along the lines of the standard derivation of the path integral representation for quantum mechanics, see \cite{Faddeev-Slavnov}.

\subsection{Representation of $Cl(\g)$, complex polarization of $\g$}
\label{sec: polarization of g}
The Clifford algebra $Cl(\g)$ admits a representation $\rho$ on the space of polynomials of $m$ odd variables $\Fun(\CC^{0|m})\cong \CC[\eta^1,\ldots,\eta^m]$. This representation is defined on  generators of $Cl(\g)$  as
\be \rho:\;\hat\psi^{a}\mapsto \left\{\begin{array}{ll}
\frac{1}{\sqrt 2}\left(\eta^p+\hbar\frac{\dd}{\dd\eta^p}\right) & \mbox{ if }a=2p-1, \\
\frac{i}{\sqrt 2}\left(\eta^p-\hbar\frac{\dd}{\dd\eta^p}\right) & \mbox{ if }a=2p.
\end{array}\right. \label{Cl rep}\ee
In fact, $\rho: Cl(\g)\ra \End(\Fun(\CC^{0|m}))\cong \End(\CC^{2^{m-1}|2^{m-1}})$ is an isomorphism of super-algebras. There is a natural identification
$$\phi:\quad \End(\Fun(\CC^{0|m}))\xra{\sim}\Fun(\CC^{0|m}\oplus \CC^{0|m})\cong \CC[\eta^1,\ldots,\eta^m,\bar\eta^1,\ldots,\bar\eta^m] .$$
This  is not an algebra morphism with respect to the standard algebra structure on polynomials. Instead, it takes the product of endomorphisms into the
convolution; see \cite{Berezin}, \cite{Faddeev-Slavnov}). We are interested in the composition $\Phi=\phi\circ\rho:\, Cl(\g)\ra \Fun(\CC^{0|m}\oplus \CC^{0|m})$ which maps
\be \Phi:\qquad \left\{\begin{array}{lll} \hat 1 &\mapsto& e^{\frac{1}{\hbar}\sum_q \eta^q\bar\eta^q}\\
\hat\psi^{2p-1}&\mapsto &\frac{1}{\sqrt{2}}(\eta^p+\bar\eta^p) e^{\frac{1}{\hbar}\sum_q \eta^q\bar\eta^q} \\
\hat\psi^{2p}&\mapsto &\frac{i}{\sqrt{2}}(\eta^p-\bar\eta^p) e^{\frac{1}{\hbar}\sum_q \eta^q\bar\eta^q} \end{array}\right. \label{Phi}\ee
and sends the product in $Cl(\g)$ into the convolution
\begin{multline}
\Phi(\hat\alpha\cdot\hat\beta)(\eta_2,\bar\eta_1) =\\
=\hbar^{m}\int\prod_p (D\eta_1^p D\bar\eta_2^p)\; \Phi(\hat\alpha)(\eta_2,\bar\eta_2)\cdot e^{\frac{1}{\hbar}\sum_q \bar\eta_2^q\eta_1^q}\cdot \Phi(\hat\beta)(\eta_1,\bar\eta_1) .
\label{Phi on product}
\end{multline}
Formula (\ref{Phi on product}) is a key point in reconstructing the path integral from the operator formalism. Another useful identity  is as follows,
\be \Str_{Cl(\g)}(\hat\alpha)=\hbar^{m}\int\prod_p (D\eta^p D\bar\eta^p)\; e^{\frac{1}{\hbar}\sum_q \bar\eta^q\eta^q}\cdot \Phi(\hat\alpha)(\eta,\bar\eta) . \label{Str_Cl via eta}\ee

More generally, a representation of type (\ref{Cl rep}) is associated to a choice of a linear complex structure $J$ on $\g$ compatible with the pairing:
$$J: \g\ra \g,\qquad J^2=-\id_\g, \qquad (Ja,b)=-(a,Jb)\;\;\mbox{for}\;\; a,b\in \g .$$
It induces the splitting of the complexified Lie algebra $\g_\CC=\CC\otimes\g$ into ``holomorphic'' and ``anti-holomorphic'' subspaces:
\be \g_\CC=\h\oplus\bar\h , \label{g splitting}\ee
where $J$ acts on $\h,\bar\h$ by multiplication by $+i$ and $-i$, respectively. (Note that $\h$ and $\bar \h$ are complex subspaces of $\g_\CC$ with respect to the standard complex structure; bar in $\bar\h$ does not mean conjugation.) The subspaces $\h$ are $\bar\h$ are Lagrangian with respect to the pairing $(,)$. The complex Lagrangian polarization (\ref{g splitting}) induces a polarization for the parity-reversed Lie algebra
\be \Pi\g=\Pi\h\oplus \Pi\bar\h . \label{g odd splitting}\ee
We denote coordinates on $\Pi\h$ by $\eta^1,\ldots, \eta^m$ and coordinates on $\Pi\bar\h$ by $\bar\eta^1,\ldots,\bar\eta^m$.
The representation $\rho: Cl(\g)=\widehat{\Fun(\Pi\g)}\ra \End(\Fun(\Pi\h))$ sends quantized holomorphic coordinates to multiplication operators and quantized anti-holomorphic coordinates to partial derivatives:
$$\rho:\left\{\begin{array}{l}\hat\eta^p\mapsto \eta^p\cdot \, ,\\ \hat{\bar\eta}^p\mapsto \hbar\frac{\dd}{\dd \eta^p} \, . \end{array}\right.$$
Morphism (\ref{Phi}) from $Cl(\g)$ to the convolution algebra $\Fun(\Pi\h\oplus \Pi\bar\h)$ is given on generators by
$$\Phi: \left\{\begin{array}{l}\hat\eta^p\mapsto \eta^p e^{\frac{1}{\hbar}\sum_q\eta^q\bar\eta^q}, \\
\hat{\bar\eta}^p\mapsto \bar\eta^p e^{\frac{1}{\hbar}\sum_q\eta^q\bar\eta^q} , \end{array}\right.$$
and it extends to the other elements of  $Cl(\g)$ by the convolution formula (\ref{Phi on product}).

We will use notation $\pi, \bar\pi$ for projections from $\Pi\g$ to $\Pi\h$ and $\Pi\bar\h$,  respectively.  We  denote by $\iota,\bar\iota$  embeddings of $\Pi\h$ and $\Pi\bar\h$ into $\Pi\g$.


\subsection{One-dimensional Chern-Simons in terms  of Atiyah-Segal's axioms}
\label{sec: Segal's axioms}

To a point with positive orientation we associate the vector super-space (the space of states)
$$\HH_{pt^+}=\Fun(\Pi\h)\cong \CC[\eta^1,\ldots,\eta^m], $$
and to a point with negative orientation -- the dual space
$$\HH_{pt^-}=(\HH_{pt^+})^*=\Fun(\Pi\bar\h)\cong \CC[\bar\eta^1,\ldots,\bar\eta^m].$$
To an interval $\I=[\p_1,\p_2]$ we associate the partition function
$$Z_\I^\rho:=\rho(Z_\I)\in \Fun(\Pi\g\oplus \g)\otimes \underbrace{\HH_{\p_2^+}\otimes \HH_{\p_1^-}}_{\cong\End(\HH_{pt^+})}$$
given by formula (\ref{Z_I}) in representation $\rho$ (see equation (\ref{Cl rep})).
In general, to a one-dimensional simplicial complex $\Theta$ we associate the partition function (\ref{Z_Theta}) of section \ref{sec: operator formalism, 1d simp cs} taken in representation $\rho$:
$$Z_\Theta^\rho:=\rho^{\otimes i(\Theta)}\circ Z_\Theta\in \Fun(\FF_\Theta^\bulk)\otimes (\HH_{pt^+}\otimes \HH_{pt^-})^{\otimes i(\Theta)} .$$

The one-dimensional Chern-Simons theory features three types of operations:
\begin{itemize}
\item To a disjoint union $\Theta_1\sqcup\Theta_2$ corresponds the tensor product for partition functions.
\item 
To a sewing of boundary points $\p_1^-$ and $\p_2^+$ in a simplicial complex $\Theta$ corresponds the convolution of spaces of states $\HH_{\p_2^+}$ and $\HH_{\p_1^-}$.
\item To a simplicial aggregation $r:\Theta\ra \Theta'$ corresponds a fiber BV integral $r_*$ which reduces the space of bulk fields from $\FF_\Theta^\bulk$ to $\FF_{\Theta'}^\bulk$.
\end{itemize}

In addition, $\HH_{pt^+}$ is equipped with an odd third-order differential operator
$$\delta^\rho=\rho(\hat \theta): \HH_{pt^+}\ra \HH_{pt^+},$$
and $\HH_{pt^-}$ is equipped with minus its dual $-(\delta^\rho)^*: \HH_{pt^-}\ra \HH_{pt^-}$. The partition function $Z_\Theta^\rho$ satisfies the quantum master equation
\be (\hbar \Delta_\Theta^\bulk+\hbar^{-1} \delta^\rho_\Theta)Z_\Theta^\rho=0 , \label{QME for Z^rho_Theta}\ee
where the ``boundary BV operator'' $\delta^\rho_\Theta$ is the sum over boundary points of $\Theta$ of operators $\delta^\rho$ or $-(\delta^\rho)^*$ acting on the corresponding $\HH_{pt}$ (depending on whether the orientation of $pt$ is positive or negative).

\begin{Rem}
$\delta^\rho$ is ``almost'' a coboundary operator: its square is proportional to identity:
\be (\delta^\rho)^2=-\frac{\hbar^3}{48}f^{abc}f^{abc}\cdot \id_{\HH_{pt^+}} \label{delta^2}\ee
(see \cite{KS}).
This implies that the boundary BV operator for an interval
$$ \delta_\I^\rho =\delta^\rho \otimes \id_{\HH_{pt^-}} - \id_{\HH_{pt^+}}\otimes (\delta^\rho)^*:\qquad \underbrace{\HH_{pt^+}\otimes \HH_{pt^-}}_{\cong\End(\HH_{pt^+})}\ra \underbrace{\HH_{pt^+}\otimes \HH_{pt^-}}_{\cong\End(\HH_{pt^+})}$$ squares to zero
$$(\delta_\I^\rho)^2=0$$
Cases when $\delta^{\rho}$  squares to zero ({\em i.e.} when $f^{abc}f^{abc}=0$) are quite interesting as then the reduced space of states for a point $\HH^\mr{red}_{pt^+}$ emerges (see remarks \ref{Rem: H^red}, \ref{Rem: ff=0}, \ref{Rem: BF+B^3} below).
\end{Rem}

\begin{Rem} The space of states $\HH_{pt^+}$ can be viewed as a geometric quantization of the classical phase super-space $\Pi\g$ (viewed as an odd K\"ahler manifold).
The operator $\delta^\rho$ is the quantization of the Maurer-Cartan element $\theta$ (\ref{theta}); the operator $\hbar^{-1} \delta^\rho_\I$ is the quantization of the Hamiltonian vector field $\{\theta,\bt\}$ on $\Pi\g$.
\end{Rem}

\begin{Rem}
Topological quantum mechanics (TQM) in the sense of A. Losev \cite{Losev TQM} assigns to an interval a manifold $Geom$ (the ``space of geometric data'') and to a point --- a vector superspace $\HH$ endowed with an odd coboundary operator $Q$. The evolution operator $U$ for an interval is a differential form on $Geom$ with values in $\End(\HH)$ and has to satisfy the ``homotopy topologicity'' equation
\be (d+\ad_Q)\; U=0 , \label{TQM eq1}\ee
where $d$ is the de Rham operator on $Geom$. A standard class of examples of TQMs comes from choosing $Geom = \mathbb{R}_{>0}$ (with coordinate $t>0$) and setting
\be U(t,dt)=e^{[Q,G]\,t+dt\; G}= e^{(d+\ad_Q)\circ (tG)} \label{TQM eq2}\ee
where $G$ is an odd operator on $\HH$.
For instance,  for the Hodge TQM \cite{Losev TQM} on a Riemannian manifold $M$, one sets $\HH=\Omega^\bt(M)$, $Q=d_M$ and $G=d^*_M$ --- the Hodge operator on forms on $M$. For the Morse TQM \cite{Witten Morse}, \cite{FLN}, one takes the same $\HH$ and $Q$, but now $G=\iota_v$ is the substitution of the gradient vector field.
The one-dimensional Chern-Simons theory on an interval can be viewed as a TQM: here $Geom=\g$ (with coordinates $A^a$), $\HH=\Fun(\Pi\h)$, $Q=\hbar^{-1}\delta^\rho$. The odd Fourier transform in variable $\tilde\psi$ of the partition function for an interval (\ref{Z_I}) is
\be U(A,\lambda)=\rho\left(e^{-\frac{1}{2\hbar}f^{abc}\hat\psi^a A^b \hat\psi^c + \lambda^a \hat\psi^a}\right)=
e^{(d+\hbar^{-1}\ad_{\delta^\rho})\circ\, \hbar^{-1}\rho\left(A^a \hat\psi^a\right)} \label{TQM 1d CS}\ee
where $d=\hbar\, \lambda^a \frac{\dd}{\dd A^a}$ is the de Rham operator on $Geom$. Note that the expression (\ref{TQM 1d CS}) is similar to (\ref{TQM eq2}) where we make a substitution $t G \mapsto \hbar^{-1}\rho(A^a \hat\psi^a)$. The quantum master equation (\ref{QME for Z_I}) is equivalent to
\be (d+\hbar^{-1}\ad_{\delta^\rho})\; U(A,\lambda)=0 \ee
which is exactly the ``homotopy topologicity'' equation (\ref{TQM eq1}). The peculiarity of the one-dimensional Chern-Simons theory viewed as a TQM is that $\delta^\rho$ is not necessarily a coboundary operator on $\HH$.
%
\end{Rem}

\subsection{Integrating out the bulk fields}
In section \ref{sec: aggregations}, we discussed simplicial aggregations which reduce the space of bulk fields of the 1-dimensional Chern-Simons theory $\FF^\bulk_\Theta \ra \FF^\bulk_{\Theta'}$ according to combinatorial moves applied to the triangulation $\Theta\ra \Theta'$. It is interesting to consider the ``ultimate aggregation'' --- integrating out the bulk fields completely. This procedure should yield the partition function in the sense of Atiyah-Segal ({\em i.e.} without bulk fields). We will denote it by $Z^\circ$.

For an interval, we have
\be (i\hbar)^m\int D\tilde\psi\; Z_\I(\tilde\psi,A)=e^{-\frac{1}{2\hbar}f^{abc}\hat\psi^a A^b \hat\psi^c} . \label{integrating out F^bulk}\ee
We view (\ref{integrating out F^bulk}) as a BV integral over the Lagrangian subspace
\be \LL_A=\{\tilde\psi+A| \; \tilde\psi \;\mbox{is free} , A\;\mbox{fixed} \} \subset \FF^\bulk_\I  . \label{L_A}\ee
This subspace depends on the value of $A$, and integral (\ref{integrating out F^bulk}) also depends on $A$. However, this dependence is $\ad_{\hat \theta}$-exact:
\be e^{-\frac{1}{2\hbar}f^{abc}\hat\psi^a (A+\delta A)^b \hat\psi^c} - e^{-\frac{1}{2\hbar}f^{abc}\hat\psi^a A^b \hat\psi^c}= \frac{1}{\hbar}\; [\hat \theta, e^{-\frac{1}{2\hbar}f^{abc}\hat\psi^a A^b \hat\psi^c+ \frac{1}{\hbar}\, \delta A^a\, \hat\psi^a}]_{Cl(\g)}+ \mc{O}((\delta A)^2) \label{integrating out F^bulk 1}\ee
(this can be checked analogously to the proof of lemma \ref{lemma: QME}). Therefore, we should understand the partition function $Z^\circ_\I$ as an element of cohomology
of the operator $\ad_{\hat \theta}$ (the fact that (\ref{integrating out F^bulk}) is $\ad_{\hat \theta}$-closed is an immediate consequence of the quantum master equation (\ref{QME for Z_I})). More exactly, $Z^\circ_\I$ is the class of Clifford unit $\hat 1$ in $\ad_{\hat \theta}$-cohomology:
\be Z^\circ_\I=[\hat 1]\in H_{\ad_{\hat \theta}}(Cl(\g))  \label{Z_I^circ}\ee
Equivalently, in terms of representation $\rho$, we have
\be \rho(Z^\circ_\I)=[\id_{\HH_{pt^{+}}}]\in H_{\delta_\I^\rho}(\End(\HH_{pt^{+}})) . \label{Z_I^circ rho}\ee

\begin{Rem}
If the contraction of structure constants $f^{abc}f^{abc}$ for $\g$ is nonzero,
the cohomology class (\ref{Z_I^circ}), (\ref{Z_I^circ rho}) vanishes since
$$\hat 1= -\frac{24\hbar^{-3}}{f^{abc}f^{abc}}\;\ad_{\hat \theta}\hat \theta$$
In fact, whole cohomology group $H_{\ad_{\hat \theta}}\cong H_{\delta_\I^\rho}$ vanishes since very  $\ad_{\hat \theta}$-cocycle $\hat\alpha \in Cl(\g)$  is automatically exact:
$$\hat\alpha=-\frac{24\hbar^{-3}}{f^{abc}f^{abc}}\;\ad_{\hat \theta} (\hat \theta\cdot \hat\alpha) .$$
\end{Rem}

\begin{Rem} \label{Rem: H^red}
If $f^{abc}f^{abc}=0$,  we can define the reduced space of states for a point as the $\delta^{\rho}$-cohomology:
$$\HH^\mr{red}_{pt^+}:= H_{\delta^\rho}(\HH_{pt^+}), \qquad \HH^\mr{red}_{pt^-}:= H_{-(\delta^\rho)^*}(\HH_{pt^-})= (\HH^\mr{red}_{pt^+})^*$$
By K\"unneth formula, we have
\be H_{\delta_\I^\rho}(\End(\HH_{pt^{+}})) \cong \HH^\mr{red}_{pt^+}\otimes \HH^\mr{red}_{pt^-} . \label{H of End(H_pt)}\ee
The partition function (\ref{Z_I^circ rho}) is then represented by the identity operator
$$\rho(Z_\I^\circ):\quad \HH^\mr{red}_{pt^+}\xrightarrow{\id} \HH^\mr{red}_{pt^+} .$$
\end{Rem}

For a circle, we can obtain the partition function $Z_{\s^1}^\circ$ (which is just a number) either as a Clifford super-trace of (\ref{integrating out F^bulk}) or as a BV integral of the effective action (\ref{W on coh of circle}) over the Lagrangian subspace (\ref{L_A}). Either way, we have
\be Z_{\s^1}^\circ=0 \ee
due to non-saturation of fermionic modes either in Clifford super-trace or in the Berezin integral over $\tilde\psi$.

\subsection{From operator formalism to path integral.}
\label{sec: back to PI}

\subsubsection{Abelian one-dimensional Chern-Simons theory}
\label{sec: back to path integral, abelian case}
The one-dimensional abelian Chern-Simons theory associates to an interval $\I$ the unit of $Cl(\g)$ (here $\g$ can be viewed  as a Euclidean vector space; the Lie algebra structure is irrelevant). The path integral arises upon applying the map (\ref{Phi}) to this trivial partition function:
\begin{multline}
\Phi(\hat 1)(\eta_{out},\bar\eta_{in})=\Phi(\underbrace{\hat 1\cdot\hat 1\cdots \hat 1}_{N})(\eta_{out},\bar\eta_{in})=\\
=\int \left(\prod_{k=1}^{N-1}\hbar^m D\eta_k D\bar\eta_{k+1}\right)\cdot \\
\cdot \exp\frac{1}{\hbar}\left(\langle\eta_{out},\bar\eta_N\rangle+\langle\bar\eta_N,\eta_{N-1}\rangle+ \langle\eta_{N-1},\bar\eta_{N-1}\rangle+\cdots +\langle\eta_2,\bar\eta_2\rangle+\langle\bar\eta_2,\eta_1\rangle+\langle\eta_1,\bar\eta_{in}\rangle\right)= \\
=\int \left(\prod_{k=1}^{N-1}\hbar^m D\eta_k D\bar\eta_{k+1}\right)\cdot \exp\frac{1}{\hbar}\left(\langle\eta_1,\bar\eta_{in}\rangle+\sum_{k=2}^N \langle \eta_k-\eta_{k-1},\bar\eta_k\rangle\right) .
\label{Phi(1) integral}
\end{multline}
For convenience, we set $\bar\eta_1:=\bar\eta_{in}$, $\eta_N:=\eta_{out}$ and introduced  a notation
$$\langle \eta,\bar\eta \rangle:=\sum_q \eta^q \bar\eta^q  .$$
In the exponential of (\ref{Phi(1) integral}), $N$ terms of type $\langle\eta_k,\bar\eta_k\rangle$  correspond to Clifford units, and $N-1$ terms of type $\langle\bar\eta_{k+1},\eta_k\rangle$ correspond to convolutions kernels as in (\ref{Phi on product}); the symbol $D\eta_k D\bar\eta_{k+1}$ is defined as $\prod_p(D\eta_k^p D\bar\eta_{k+1}^p)$. Expression (\ref{Phi(1) integral}) corresponds to triangulating an interval by  $N$ smaller intervals; the terms in the exponential correspond to 0- and 1-simplices of this triangulation. In the limit $N\ra\infty$, one formally writes (\ref{Phi(1) integral}) as a path integral over paths $\eta(\tau)$, $\bar\eta(\tau)$ with $\eta$ at the right end-point and $\bar\eta$ at left end-point of $\I$ fixed by the boundary conditions.
\begin{lemma}
\be \Phi(\hat 1)(\eta_{out},\bar\eta_{in})=\int_{\bar\eta(0)=\bar\eta_{in},\;\eta(1)=\eta_{out}}\DD\eta\DD\bar\eta\cdot \exp\frac{1}{\hbar} \left(\langle \eta(0), \bar\eta(0) \rangle+\int_\I \langle d\eta, \bar\eta \rangle\right) \label{Phi(1) path integral}\ee
\end{lemma}

A perturbative computation of the path integral (\ref{Phi(1) path integral}) is trivial: the integral is given by the contribution of the critical point
$$\eta(\tau)=\eta_{out},\quad \bar\eta(\tau)=\bar\eta_{in}\quad \mbox{for all }\; \tau\in [0,1]$$
and yields
$$\Phi(\hat 1)(\eta_{out},\bar\eta_{in}) =e^{\frac{1}{\hbar}\langle \eta_{out}, \bar\eta_{in} \rangle} .$$

It is instructive to write the integral (\ref{Phi(1) path integral}) in terms of the field $\psi$ instead of fields $\eta,\bar\eta$. For simplicity, we first choose a
 complex polarization (as in (\ref{Cl rep}))
\be\left\{\begin{array}{lll}\psi^{2p-1}&=&\frac{1}{\sqrt{2}}(\eta^p+\bar\eta^p) \\ \psi^{2p}&=&\frac{i}{\sqrt{2}}(\eta^p-\bar\eta^p) \end{array}\right. \qquad \Leftrightarrow \qquad \left\{\begin{array}{lll}\eta^p&=&\frac{1}{\sqrt{2}}(\psi^{2p-1}-i\psi^{2p}) \\ \bar\eta^p&=&\frac{1}{\sqrt{2}}(\psi^{2p-1}+i\psi^{2p}) \end{array}\right. \label{psi and eta} \ee
(this corresponds to the complex structure $J$ on $\g$ which assigns $\psi^{2p-1}$ as ``real'' coordinates and $\psi^{2p}$ as ``imaginary'' coordinates on $\Pi\g$). We have,
\begin{eqnarray*}\sum_p\eta_k^p \bar\eta_k^p &=& \sum_p i \psi^{2p-1}_k \psi^{2p}_k, \\
\sum_p \bar\eta_{k+1}^p \eta_k^p &=& \sum_p\frac{\psi_{k+1}^{2p-1}\psi_k^{2p-1}+\psi_{k+1}^{2p}\psi_k^{2p}}{2}-\sum_p i\frac{\psi_{k+1}^{2p-1}\psi_k^{2p}-\psi_{k+1}^{2p}\psi_k^{2p-1}}{2} .
\end{eqnarray*}
Substituting these expressions into the integral representation (\ref{Phi(1) integral}) for $\Phi(1)$, we obtain
\begin{multline} \label{Phi(1) psi integral}
\Phi(\hat 1)(\eta_{out},\bar\eta_{in})=\int \left(\hbar^{m/2} D \eta_1\right) \left(\prod_{k=2}^{N-1} (i\hbar)^m D\psi_k\right) \left(\hbar^{m/2} D \bar\eta_N\right)\cdot \\
\cdot\exp\frac{1}{\hbar} \left(\sum_{k=1}^{N-1}\frac{1}{2}(\psi_{k+1},\psi_k)+\sum_p\sum_{k=1}^{N-1} \frac{i}{2}(\psi_{k+1}^{2p-1}-\psi_k^{2p-1})(\psi_{k+1}^{2p}-\psi_k^{2p})+\right. \\
\left.+\sum_p\frac{i}{2}\psi_1^{2p-1}\psi_1^{2p}+ \sum_p\frac{i}{2}\psi_N^{2p-1}\psi_N^{2p}\right) .
\end{multline}
Here $D\psi_k:=\ola\prod_a D\psi_k^a$ is the Berezin measure on $\Pi\g$; the variable $\psi_1$ is constructed by formulae (\ref{psi and eta}) from the integration variable $\eta_1$ and the boundary value $\bar\eta_1:=\bar\eta_{in}$, and $\psi_N$ is constructed from the integration variable $\bar\eta_N$ and the boundary value $\eta_N:=\eta_{out}$.

\begin{Rem} We think of integral (\ref{Phi(1) integral}) as corresponding to cutting the interval $\I=[\p_{in},\p_{out}]$ into $N$ intervals $[\p_{in},\p_2]\cup [\p_2,\p_3]\cup\cdots \cup [\p_{N},\p_{out}]$. Integration variables $\eta_{k},\bar\eta_{k+1}$ are associated to the point $\p_{k+1}$ (more specifically, to the right end of interval $[\p_{k},\p_{k+1}]$ and to the left end of the interval $[\p_{k+1},\p_{k+2}]$, respectively); the boundary value $\bar\eta_{in}$ corresponds to the point $\p_{in}$, the boundary value $\eta_{out}$ --- to the point $\p_{out}$. However, variables $\psi_k$ are linear combinations of $\eta_k$ and $\bar\eta_k$. Thus, they are not associated to any single point, but rather to a pair of neighboring points $(\p_{k},\p_{k+1})$.
\end{Rem}

Again, we formally write the limit $N\ra\infty$ of the integral (\ref{Phi(1) psi integral}) as a path integral over paths $\psi: \I\ra \Pi\g$ with a fixed anti-holomorphic projection of $\psi$ at the  right end-point of the interval and a fixed holomorphic projection at the left end-point:
\begin{lemma}
The path integral expression for the partition function of the abelian Chern-Simons theory on an interval is given by
\begin{multline}
\Phi(\hat 1)(\eta_{out},\bar\eta_{in})= \int_{\bar\pi(\psi(0))=\bar\eta_{in},\; \pi(\psi(1))=\eta_{out}} \DD\psi\cdot \\
\cdot \exp \frac{1}{\hbar}\left(\sum_p \frac{i}{2} \psi^{2p-1}(0)\psi^{2p}(0)+\int_\I \frac{1}{2}(\psi,d\psi) + \sum_p \frac{i}{2} \psi^{2p-1}(1)\psi^{2p}(1)\right) .
\label{Phi(1) psi path integral}
\end{multline}
\end{lemma}
The exact meaning of the conditional measure on paths in (\ref{Phi(1) psi path integral}) is the formal $N\ra \infty$ limit of the measure in (\ref{Phi(1) psi integral}). The second term in the exponential in (\ref{Phi(1) psi integral}) does not contribute to the limit $N\ra\infty$: once we assume that $\psi_k$ are values of a differentiable path $\psi(\tau)$ at times $\tau=k/N$, the contribution of this term becomes of order $\mc{O}(1/N)$.

For a general complex structure $J$ on $\g$, path integral (\ref{Phi(1) psi path integral}) becomes
\begin{multline*}
\Phi(\hat 1)(\eta_{out},\bar\eta_{in})=\\
= \int_{\bar\pi(\psi(0))=\bar\eta_{in},\; \pi(\psi(1))=\eta_{out}} \DD\psi
\cdot \exp \frac{1}{\hbar}\left(\frac{i}{4} (\psi(0),J\psi(0))+\int_\I \frac{1}{2}(\psi,d\psi) + \frac{i}{4} (\psi(1),J\psi(1))\right) .
\end{multline*}

\subsubsection{Path integral for the non-abelian one-dimensional Chern-Simons theory in the cyclic Whitney gauge. End of proof of theorem \ref{thm: claim}.}
\label{sec: back to path integral, non-abelian case}

To obtain a path integral representation for the partition function of the one-dimensional Chern-Simons theory on an interval (\ref{Z_I}) we use the same strategy as in section \ref{sec: back to path integral, abelian case}: we cut the interval into $N$ smaller intervals and then apply the map $\Phi$ (\ref{Phi}). The new point here is that  for small intervals we have to use the ``heat kernel'' approximation which gives an exact result only in the limit $N\ra\infty$.

Applying $\Phi$ to $Z_\I$ (\ref{Z_I}), we have
\begin{multline} \label{Phi(Z_I)}
\Phi(Z_\I)(\eta_{out},\bar\eta_{in})=  \int (i\hbar)^{-m} D\lambda\cdot e^{-(\lambda,\tilde\psi)}\cdot \Phi\left(\exp\left(-\frac{1}{2\hbar}(\hat\psi,[A,\hat\psi])+(\lambda,\hat\psi)\right)\right) \eta_{out},\bar\eta_{in})=\\
=\int (i\hbar)^{-m} D\lambda\cdot e^{-(\lambda,\tilde\psi)}\cdot \Phi\left(\left(\exp\left(-\frac{1}{2\hbar N}(\hat\psi,[A,\hat\psi])+\frac{1}{N}(\lambda,\hat\psi)\right)\right)^N\right)(\eta_{out},\bar\eta_{in})=\\
=\int (i\hbar)^{-m} D\lambda\cdot e^{-(\lambda,\tilde\psi)}\int \left(\prod_{k=1}^{N-1}\hbar^m D\eta_k D\bar\eta_{k+1}\right)\cdot \\
\cdot \Phi\left(\exp\left(-\frac{1}{2\hbar N}(\hat\psi,[A,\hat\psi])+\frac{1}{N}(\lambda,\hat\psi)\right)\right)(\eta_{out},\bar\eta_{N})\cdot e^{\frac{1}{\hbar}\langle\bar\eta_N,\eta_{N-1}\rangle}\cdots \\
\cdots  e^{\frac{1}{\hbar}\langle\bar\eta_2,\eta_{1}\rangle}\cdot \Phi\left(\exp\left(-\frac{1}{2\hbar N}(\hat\psi,[A,\hat\psi])+\frac{1}{N}(\lambda,\hat\psi)\right)\right)(\eta_1,\bar\eta_{in}) .
\end{multline}
Next, we need to evaluate the partition function for a small interval in the limit $N\ra \infty$:
\begin{multline} \label{heat kernel}
\Phi\left(\exp\left(-\frac{1}{2\hbar N}(\hat\psi,[A,\hat\psi])+\frac{1}{N}(\lambda,\hat\psi)\right)\right)(\eta,\bar\eta)=\\
=\Phi\left(\hat 1-\frac{1}{2\hbar N}(\hat\psi,[A,\hat\psi])+\frac{1}{N}(\lambda,\hat\psi)+\mc{O}\left(\frac{1}{N^2}\right)\right)(\eta,\bar\eta)=\\
= e^{\frac{1}{\hbar}\langle\eta,\bar\eta\rangle} \left(1-\frac{1}{2\hbar N}(\psi,[A,\psi])+\frac{1}{N}(\lambda,\psi)+\frac{i}{4N}\tr\left( J\cdot \ad_A\right)+\mc{O}\left(\frac{1}{N^2}\right)\right) .
\end{multline}
Here $\psi$ is a linear combination of $\eta,\bar\eta$, prescribed by the choice of a complex structure $J$ (e.g. (\ref{psi and eta})); the term with a trace appeared due to the following identity:
$$\Phi(\hat\psi^a\hat\psi^b)(\eta,\bar\eta)=e^{\frac{1}{\hbar}\langle\eta,\bar\eta\rangle} \left(\psi^a\psi^b+\frac{\hbar}{2}\delta^{ab}+\frac{i\hbar}{2}J^{ab}\right) .$$
Here the third term generates the trace term in (\ref{heat kernel}). Substituting the ``heat kernel'' asymptotics (\ref{heat kernel}) into (\ref{Phi(Z_I)}), we get
\begin{multline} \label{Phi(Z_I) approx}
\Phi(Z_\I)(\eta_{out},\bar\eta_{in})=\\
=\int \left(\prod_{k=1}^{N-1}\hbar^m D\eta_k D\bar\eta_{k+1}\right)  \int (i\hbar)^{-m} D\lambda\cdot e^{(\lambda,\frac{1}{N}\sum_{k=1}^N\psi_k-\tilde\psi)}\cdot \\
\cdot e^{\frac{i}{4}\tr(J\cdot \ad_{A})} \cdot e^{\frac{1}{\hbar}(\langle\eta_{out},\bar\eta_N\rangle+\langle\bar\eta_N,\eta_{N-1}\rangle \cdots +\langle\bar\eta_2,\eta_1\rangle +\langle\eta_1,\bar\eta_{in}\rangle)}\cdot
e^{-\frac{1}{2\hbar N}\sum_{k=1}^N (\psi_k,[A,\psi_k])}+ \mc{O}\left(\frac{1}{N}\right) .
\end{multline}
Taking the limit $N\ra\infty$, we obtain the following.
\begin{proposition}
The Chern-Simons partition function for an interval is given by the  path integral:
\begin{multline} \label{Phi(Z_I) path integral}
\Phi(Z_\I)(\eta_{out},\bar\eta_{in})=e^{\frac{i}{4}\tr(J\cdot \ad_{A})}\int_{\bar\pi(\psi(0))=\bar\eta_{in},\; \pi(\psi(1))=\eta_{out}, \; \int_\I d\tau \psi(\tau)=\tilde\psi} \DD\psi\cdot \\
\cdot\exp{\frac{1}{\hbar}\left(\int_\I \frac{1}{2}(\psi,(d+d\tau\cdot\ad_A)\psi)+\frac{i}{4} (\psi(0),J\psi(0))+\frac{i}{4} (\psi(1),J\psi(1))\right)}
\end{multline}
The conditional measure on paths $\psi(\tau)$ with a fixed holomorphic projection at $\tau=1$, a fixed anti-holomorphic projection at $\tau=0$ and with a fixed integral over $\tau$ in (\ref{Phi(Z_I) path integral}) is the $N\ra\infty$ limit of the measure in (\ref{Phi(Z_I) approx}).
\end{proposition}
Applying concatenation formulae (\ref{Phi on product}), (\ref{Str_Cl via eta}) to (\ref{Phi(Z_I) path integral}), we obtain a path integral representation of the Chern-Simons partition function for one-dimensional simplicial complexes:
\begin{corollary} For a triangulated interval $\Theta=[\p_{in},\p_2]\cup [\p_2,\p_3]\cup \cdots \cup [\p_{n},\p_{out}]$ we have
\begin{multline} \Phi(Z_\Theta)(\eta_{out},\bar\eta_{in})=e^{\sum_{k=1}^n\frac{i}{4}\tr(J\cdot \ad_{A_k})} \int_{\bar\pi(\psi(\p_{in}))=\bar\eta_{in},\; \pi(\psi(\p_{out}))=\eta_{out}, \; \int_{\p_{k}}^{\p_{k+1}} d\tau \psi(\tau)=\tilde\psi_k} \DD\psi\cdot \\
\cdot\exp{\frac{1}{\hbar}\left(\int_\I \frac{1}{2}(\psi,(d+d\tau\cdot\ad_A)\psi)+\frac{i}{4} (\psi(\p_{in}),J\psi(\p_{in}))+\frac{i}{4} (\psi(\p_{out}),J\psi(\p_{out}))\right)} .
\label{Z of Theta path integral}
\end{multline}
For a triangulated circle $\Xi_n=[\p_1,\p_2]\cup \cdots \cup [\p_{n-1},\p_n]\cup [\p_n,\p_1]$, we obtain:
\be \label{Z of X_n path integral}
Z_{\Xi_n}
=e^{\sum_{k=1}^n\frac{i}{4}\tr(J\cdot \ad_{A_k})} \int_{\int_{\p_{k}}^{\p_{k+1}} d\tau \psi(\tau)=\tilde\psi_k} \DD\psi
\cdot e^{\frac{1}{2\hbar}\int_\I (\psi,(d+d\tau\cdot\ad_A)\psi)} .
\ee
\end{corollary}

\begin{Rem}
Expression (\ref{Z of X_n path integral}) returns us to the ``na\"ive'' path-integral (\ref{S_Xi functional integral}) for the simplicial Chern-Simons on a circle, up to a somewhat puzzling factor $e^{\sum_k\frac{i}{4}\tr(J\cdot \ad_{A_k})}$. The explanation is as follows: the path integral in (\ref{Z of X_n path integral}) was obtained from the path integral with boundaries (\ref{Z of Theta path integral}) by concatenation formula (\ref{Str_Cl via eta}). Hence, it is  secretly using the normal ordering prescribed by the choice of a complex structure $J$ on $\g$ (which dictates the regularization for the one-loop determinant in (\ref{Z of X_n path integral})). This implicit dependence on $J$ is exactly cancelled by the factor $e^{\sum_k\frac{i}{4}\tr(J\cdot \ad_{A_k})}$ (indeed, we know that the left hand side of (\ref{Z of X_n path integral}) is defined in terms of the Clifford algebra $Cl(\g)$ and therefore cannot possibly depend on $J$).
For the na\"ive path integral (\ref{S_Xi functional integral}), we implicitly assumed the symmetric normal ordering by making a regularization (\ref{theta(0)}) in our computation of the one-loop determinant (the important point is that $\theta(0)$ is a number, and not a matrix).
%
\end{Rem}
Path integral representation (\ref{Z of X_n path integral}) returns us to perturbative computation of section \ref{sec: simplicial CS on circle 1.3} and thus finishes the proof of theorem \ref{thm: claim}.

\begin{Rem} It is easy to compute (\ref{Phi(Z_I) path integral}) in the case of $A=0$:
\be \Phi(Z_\I|_{A=0})(\eta_{out},\bar\eta_{in})=2^{-m}\,e^{\frac{1}{\hbar}\left(\langle\eta_{out},\bar\eta_{in} \rangle-2\langle\eta_{out}-\tilde\eta, \bar\eta_{in}-\bar{\tilde\eta}\rangle\right)} ,\label{Phi(Z_I) for A=0}\ee
where $\tilde\eta, \bar{\tilde\eta}$ are holomorphic and anti-holomorphic components of the bulk field $\tilde\psi$. We can also write (\ref{Phi(Z_I) for A=0}) as
$$\Phi(Z_\I|_{A=0})(\eta_{out},\bar\eta_{in})=2^{-m}\,e^{\frac{1}{\hbar}\left(\frac{i}{2}(\psi_{bd},J\psi_{bd})-i (\psi_{bd}-\tilde\psi,J(\psi_{bd}-\tilde\psi))\right)} ,$$
where $\psi_{bd}=\iota(\eta_{out})+\bar\iota(\bar\eta_{in})$ is the linear combination of boundary fields $\eta_{out},\bar\eta_{in}$.
\end{Rem}

\subsection{Simplicial action on an interval}
\label{sec: simp action on interval}
\begin{proposition}
The path integral for the Chern-Simons partition function on an interval (\ref{Phi(Z_I) path integral}) is given by
\begin{multline}
\Phi(Z_\I(\tilde\psi,A))(\eta_{out},\bar\eta_{in})= {\det}^{1/2}_\g\left(\frac{\sinh\frac{\ad_A}{2}}{\frac{\ad_A}{2}}\right)\cdot {\det}^{-1/2}_\g M(\ad_A)\cdot \\
\cdot\exp\frac{
1}{\hbar}\left(\langle\eta_{out},\bar\eta_{in}\rangle- \frac{1}{2}(\tilde\psi,\ad_A \tilde\psi)+\frac{1}{2} \left(\begin{array}{ll} \tilde\eta-\eta_{out} & \bar{\tilde\eta}-\bar\eta_{in} \end{array}\right)\cdot M(\ad_A)\cdot\left(\begin{array}{l} \tilde\eta-\eta_{out} \\ \bar{\tilde\eta}-\bar\eta_{in} \end{array}\right) \right) ,
\label{Phi(Z_I) quasi-classic}
\end{multline}
where the bilinear form $M(\ad_A)$ in basis $(\eta,\bar\eta)$ is represented by the block matrix
\be M(\ad_A)=\left(\begin{array}{cc} R_{-+}R_{++}^{-1} & -1-R_{--}+R_{-+}R_{++}^{-1}R_{+-} \\
1+R_{++}^{-1} & R_{++}^{-1} R_{+-}
\end{array}\right) . \label{M}\ee
Here symbols $R_{\pm \pm}$ stand for blocks of $R(\ad_A)$ (defined by formula (\ref{R(A)})) in the basis $(\eta,\bar\eta)$:
\be R(\ad_A)=\left(\begin{array}{ll}R_{++} & R_{+-} \\ R_{-+} & R_{--}\end{array}\right) . \label{R blocks}  \ee
\end{proposition}

\textit{Proof.} The path integral (\ref{Phi(Z_I) path integral}) is Gaussian with a critical point $\psi^{cr}(\tau)$ being the solution of
\be (d+d\tau\, \ad_A)\psi^{cr}=const \label{psi^cr eq}\ee
subject to conditions
\begin{eqnarray}
\int d\tau\, \psi^{cr}(\tau)&=&\tilde\psi , \label{psi^cr cond1}\\
\pi(\psi^{cr}(1))&=&\eta_{out} , \label{psi^cr cond2}\\
\bar\pi(\psi^{cr}(0))&=&\bar\eta_{in} . \label{psi^cr cond3}
\end{eqnarray}
Equation (\ref{psi^cr eq}) together with (\ref{psi^cr cond1}) gives
\be \psi^{cr}(\tau)=\tilde\psi+F(\ad_A,\tau)(\psi^{cr}(0)-\tilde{\psi}) , \label{psi^cr 1}\ee
where $F(\ad_A,\tau)$ is defined by (\ref{F(A)}). The value of the action (together with boundary terms) for this path is
\begin{multline}
S(\psi^{cr})+\mbox{boundary terms}= \\
=\underbrace{-\frac{1}{2}(\tilde\psi,\ad_A\tilde\psi)-\frac{1}{2}(\tilde\psi,\psi^{cr}(1)-\psi^{cr}(0))}_{\frac{1}{2}\int (\psi^{cr},d_A \psi^{cr})}+\frac{1}{2}\langle\eta^{cr}(0),\bar\eta_{in}\rangle+ \frac{1}{2}\langle\eta_{out},\bar\eta^{cr}(1)\rangle  . \label{psi^cr action}
\end{multline}
Boundary values of $\psi^{cr}$ can be found as follows: (\ref{psi^cr 1}) implies that
$$\psi^{cr}(1)=(1+R(\ad_A))\tilde\psi - R(\ad_A)\psi^{cr}(0) . $$
Solving equation this together with (\ref{psi^cr cond2}) and (\ref{psi^cr cond3}) in coordinates $(\eta,\bar\eta)$, we obtain
\be \left(\begin{array}{l}\eta_{out} \\ \bar\eta^{cr}(1)\end{array}\right)=
\left(\begin{array}{cc}1+R_{++} & R_{+-} \\ R_{-+} & 1+R_{--}\end{array}\right)\left(\begin{array}{l}\tilde\eta \\ \bar{\tilde\eta}\end{array}\right)- \left(\begin{array}{cc}R_{++} & R_{+-} \\ R_{-+} & R_{--}\end{array}\right) \left(\begin{array}{l}\eta^{cr}(0) \\ \bar\eta_{in}\end{array}\right) . \label{psi^cr 2} \ee
Here $\eta^{cr}(0)$ and $\bar\eta^{cr}(1)$ are the unknowns. Solving (\ref{psi^cr 2}), we get
\begin{eqnarray*}
\eta^{cr}(0)&=& \eta_{out}+(1+R_{++}^{-1})(\tilde\eta-\eta_{out}) + R_{++}^{-1} R_{+-} (\bar{\tilde\eta} -  \bar\eta_{in}) , \\
\bar\eta^{cr}(1)&=& \bar\eta_{in}-R_{-+}R_{++}^{-1} (\tilde\eta-\eta_{out}) - (-1-R_{--}+R_{-+}R_{++}^{-1}R_{+-})(\bar{\tilde\eta}-\bar\eta_{in}) .
\end{eqnarray*}
By substituting into (\ref{psi^cr action}), we obtain the exponential in (\ref{Phi(Z_I) quasi-classic}).

The pre-exponential in (\ref{Phi(Z_I) quasi-classic}) can be derived as follows. We know that it is a function of $A$ only (since it is a square root of a functional determinant of the operator\footnote{More exactly, the  determinant of a matrix of the bilinear form $\int (\bt,d_A \bt)$.} $d_A$  acting on functions with vanishing integral, vanishing holomorphic part at $\tau=1$ and vanishing anti-holomorphic part at $\tau=0$); let us denote it by $G(A)$. Closing the interval into a circle (by using (\ref{Str_Cl via eta})), we find
$$\int \hbar^m D\eta_{out} D\bar\eta_{in}\; e^{\frac{1}{\hbar}\langle\bar\eta_{in},\eta_{out}\rangle} \cdot \Phi(Z_\I(\tilde\psi,A))(\eta_{out},\bar\eta_{in})= {\det}^{1/2}_\g M(\ad_A)\cdot G(A)\cdot e^{-\frac{1}{2\hbar}(\tilde\psi,\ad_A\tilde\psi)} .$$
Comparing with the known result for a circle (\ref{W on coh of circle}), we obtain the pre-exponential in (\ref{Phi(Z_I) quasi-classic}). $\Box$

\begin{Rem}
In this computation, we neglected the factor of $e^{\frac{i}{4}\tr (J\cdot \ad_A)}$. If we did not, it would anyway be cancelled by the pre-exponential obtained by comparison with (\ref{W on coh of circle}) (and this result is obtained by an explicit computation in operator formalism in section \ref{sec: consistency check 1}).
\end{Rem}

We can define the simplicial Chern-Simons action for the interval $S_\I(\tilde\psi,A;\eta_{out},\bar\eta_{in})$ by
$$e^{\frac{1}{\hbar}S_\I(\tilde\psi,A;\eta_{out},\bar\eta_{in})}=\Phi(Z_\I(\tilde\psi,A))(\eta_{out},\bar\eta_{in}), $$
or explicitly
\begin{multline} \label{W_I}
S_\I(\tilde\psi,A;\eta_{out},\bar\eta_{in})=\\
=\langle\eta_{out},\bar\eta_{in}\rangle- \frac{1}{2}(\tilde\psi,\ad_A \tilde\psi)+\frac{1}{2} \left(\begin{array}{ll} \tilde\eta-\eta_{out} & \bar{\tilde\eta}-\bar\eta_{in} \end{array}\right)\cdot M(\ad_A)\cdot\left(\begin{array}{l} \tilde\eta-\eta_{out} \\ \bar{\tilde\eta}-\bar\eta_{in} \end{array}\right)+\\
+\frac{\hbar}{2}\,\tr_\g\log \left(\frac{\sinh\frac{\ad_A}{2}}{\frac{\ad_A}{2}}\right)-\frac{\hbar}{2}\,\tr_\g\log M(\ad_A) .
\end{multline}

\begin{Rem} Expansion of (\ref{W_I}) as a power series in  $A$ starts as
\begin{multline}
S_\I(\tilde\psi,A;\eta_{out},\bar\eta_{in})= \\
=\frac{i}{2}(\psi_{bd},J\psi_{bd})-i(\psi_{bd}-\tilde\psi,\, J(\psi_{bd}-\tilde\psi)) - \frac{1}{2}(\tilde\psi,\ad_A \tilde\psi) -\frac{1}{6} (\psi_{bd}-\tilde\psi,\, J\,\ad_A J \,(\psi_{bd}-\tilde\psi))+ \mc{O}(A^2)-\\
-\hbar\, m \log 2 +\hbar\, \frac{i}{12} \tr (J\,\ad_A)+ \mc{O}(\hbar A^2) ,
\end{multline}
where $\psi_{bd}=\iota(\eta_{out})+\bar\iota(\bar\eta_{in})$.
\end{Rem}

We can obtain the action for a simplicial complex $\Theta$ by gluing actions (\ref{W_I}) for individual intervals using concatenation formulae (\ref{Phi on product}), (\ref{Str_Cl via eta}). E.g. for a triangulated interval $\Theta=[\p_{in},\p_2]\cup [\p_2,\p_3]\cup \cdots \cup [\p_{n},\p_{out}]$ we have
\begin{multline}
e^{\frac{1}{\hbar}S_\Theta(\tilde\psi_1,A_1,\ldots,\tilde\psi_n,A_n;\eta_{out},\bar\eta_{in})}= \int\prod_{k=1}^{n-1}\left(\hbar^m D\eta_k D\bar\eta_{k+1}\right)\cdot \\ \cdot\exp\frac{1}{\hbar}\left(S_\I(\tilde\psi_n,A_n;\eta_{out},\bar\eta_n)+\langle\bar\eta_n,\eta_{n-1}\rangle+\cdots
+\langle\bar\eta_2,\eta_1\rangle+S_\I(\tilde\psi_1,A_1;\eta_{1},\bar\eta_{in})\right) .
\label{W_Theta}
\end{multline}
For a triangulated circle $\Xi_n=[\p_1,\p_2]\cup\cdots\cup [\p_{n-1},\p_n]\cup [\p_n,\p_1]$, we have
\begin{multline}
e^{\frac{1}{\hbar}S_{\Xi_n}(\tilde\psi_1,A_1,\ldots,\tilde\psi_n,A_n)}= \int\prod_{k=1}^{n}\left(\hbar^m D\eta_k D\bar\eta_{k+1}\right)
\cdot\exp\frac{1}{\hbar}\sum_{k=1}^n\left(S_\I(\tilde\psi_k,A_k;\eta_k,\bar\eta_k)+ \langle\bar\eta_{k+1},\eta_k\rangle\right) .
\label{W_Xi}
\end{multline}

\begin{Rem} Looking at formulae (\ref{W_Theta}), (\ref{W_Xi}), it is tempting to identify $\langle\bar\eta_{k+1},\eta_k\rangle$ as a simplicial action for the point $\p_{k+1}$.
\end{Rem}

\begin{Rem}
Formula (\ref{W_Xi}) 
explains how the simplicially non-local expression (\ref{S_Xi explicit}) is produced from a simplicially local expression (the sum of contributions of individual intervals --- the integrand in (\ref{W_Xi})). The key is  integration over boundary fields $\{\eta_k, \bar\eta_k\}$.
\end{Rem}

Let us introduce  the notation $f^{pqr}_{\pm\pm\pm}$ for  structure constants\footnote{Here we mean the structure constants of the cyclic operation $(\bt,[\bt,\bt]):\wedge^3\g\ra \RR$.} of $\g$ in the basis $(\eta,\bar\eta)$:
\be \theta=\frac{1}{6} f^{abc} \psi^a\psi^b\psi^c = \frac{1}{6}f^{pqr}_{+++}\eta^p\eta^q\eta^r + \frac{1}{2}f^{pqr}_{++-}\eta^p\eta^q\bar\eta^r+\frac{1}{2}f^{pqr}_{+--}\eta^p\bar\eta^q\bar\eta^r+ \frac{1}{6}f^{pqr}_{---}\bar\eta^p\bar\eta^q\bar\eta^r .
\label{f components}
\ee
(this is the same $\theta$ as in (\ref{theta}) rewritten in holomorphic-antiholomorphic coordinates on $\Pi\g$).
Then, the quantum master equation (\ref{QME for Z_I}) for the action (\ref{W_I}) has the following form:
\begin{multline}
\label{QME with eta}
\hbar\, \frac{\dd}{\dd\tilde\psi^a}\frac{\dd}{\dd A^a}e^{\frac{1}{\hbar}S_\I(\tilde\psi,A;\eta,\bar\eta)}
+\frac{1}{\hbar}\left( \frac{1}{6}f^{pqr}_{+++}\eta^p\eta^q\eta^r + \frac{\hbar}{2}f^{pqr}_{++-}\eta^p\eta^q \frac{\dd}{\dd\eta^r}- \frac{\hbar}{2}f^{pqq}_{++-}\eta^p+\right.
\\ \left.+\frac{\hbar^2}{2}f^{pqr}_{+--}\eta^p\frac{\dd}{\dd\eta^q}\frac{\dd}{\dd\eta^r}-
\frac{\hbar^2}{2}f^{ppr}_{+--}\frac{\dd}{\dd\eta^r}+ \frac{\hbar^3}{6}f^{pqr}_{---}\frac{\dd}{\dd\eta^p}\frac{\dd}{\dd\eta^q}\frac{\dd}{\dd\eta^r} \right) e^{\frac{1}{\hbar}S_\I(\tilde\psi,A;\eta,\bar\eta)}-\\
-e^{\frac{1}{\hbar}S_\I(\tilde\psi,A;\eta,\bar\eta)}
\frac{1}{\hbar}\left( \frac{\hbar^3}{6}f^{pqr}_{+++}\frac{\ola\dd}{\dd\bar\eta^p}\frac{\ola\dd}{\dd\bar\eta^q} \frac{\ola\dd}{\dd\bar\eta^r} + \frac{\hbar^2}{2}f^{pqr}_{++-}\frac{\ola\dd}{\dd\bar\eta^p} \frac{\ola\dd}{\dd\bar\eta^q}  \bar\eta^r- \frac{\hbar^2}{2}f^{pqq}_{++-}\frac{\ola\dd}{\dd\bar\eta^p}+\right.
\\ \left.+\frac{\hbar}{2}f^{pqr}_{+--}\frac{\ola\dd}{\dd\bar\eta^p} \bar\eta^q\bar\eta^r-
\frac{\hbar}{2}f^{ppr}_{+--}\bar\eta^r+ \frac{\hbar^3}{6}f^{pqr}_{---}\bar\eta^p\bar\eta^q \bar\eta^r \right)=0 .
\end{multline}


\begin{Rem}
The one-dimensional $BF$ theory is a special case of the one-dimensional Chern-Simons theory where the complex polarization (\ref{g splitting}) is compatible with the Lie algebra structure on $\g$. In more detail, let $\h$ in (\ref{g splitting}) be a Lie subalgebra, and let the Lie algebra structure on $\g$ be given by a semidirect product of $\h$ with its coadjoint module $\bar\h$:
$$\g=\h\ltimes \bar\h .$$
In this case, formula (\ref{W_I}) for $S_\I$ simplifies: the block $R_{+-}$ in (\ref{R blocks}) vanishes, and  the matrix $M(\ad_A)$ (\ref{M}) becomes
$$ M(\ad_A)=\left(\begin{array}{cc} R_{-+}R_{++}^{-1} & -1-R_{--} \\
1+R_{++}^{-1} & 0
\end{array}\right),\qquad {\det}_\g^{1/2} M(\ad_A)={\det}_\h (1+R_{++}^{-1}) . $$
In (\ref{f components}), only the second term on the right hand side survives:
$$\theta=\frac{1}{2}F_{pq}^r\eta^p\eta^q\bar\eta_r .$$
(Here $F_{pq}^r$ are the structure constants of $\h$; we distinguish between upper and lower indices  to emphasize that we do not assume that $\h$ comes with a pairing).
So, the quantum master equation (\ref{QME with eta}) is simplified:
\begin{multline}
\hbar\, \frac{\dd}{\dd\tilde\psi^a}\frac{\dd}{\dd A^a}e^{\frac{1}{\hbar}S_\I(\tilde\psi,A;\eta,\bar\eta)}
+\frac{1}{\hbar}\left(\frac{\hbar}{2}F_{pq}^r\eta^p\eta^q \frac{\dd}{\dd\eta_r}- \frac{\hbar}{2}F_{pq}^q\eta^p\right) e^{\frac{1}{\hbar}S_\I(\tilde\psi,A;\eta,\bar\eta)}-\\
-e^{\frac{1}{\hbar}S_\I(\tilde\psi,A;\eta,\bar\eta)}
\frac{1}{\hbar}\left(\frac{\hbar^2}{2}F_{pq}^r\frac{\ola\dd}{\dd\bar\eta^p} \frac{\ola\dd}{\dd\bar\eta^q}  \bar\eta_r- \frac{\hbar^2}{2}F_{pq}^q\frac{\ola\dd}{\dd\bar\eta^p}\right)=0 .
\end{multline}
%
%
(If in addition $\h$ is unimodular, the last terms in brackets vanish.)
Note that the result for $BF$ theory that we obtain from (\ref{W_I}) cannot be directly compared to the result in \cite{simpBF} as the choice of gauge fixing is very different\footnote{Indeed, here we fix the field $A$ to be constant on the interval, and we fix the integral $\tilde\psi$ of field $\psi$ over the interval, and the the holomorphic and anti-holomorphic projections of $\psi$ at the right and left end-points of the interval. The gauge used in \cite{simpBF}  fixes $\pi(A)$ to be constant, $\bar\pi(A)$ to be a sum of delta-functions at the ends of the interval; and it fixes the values  $\pi(\psi)$ at the ends of the interval and the integral for $\bar\pi(\psi)$. The latter gauge choice features better simplicial locality properties, but is only $\h$-equivariant.}.
\end{Rem}

\begin{Rem} \label{Rem: ff=0}
Another interesting point about the $BF$ case is that $f^{abc}f^{abc}=0$. Hence, the operator $\delta^\rho: \HH_{pt^+}\ra \HH_{pt^+}$ becomes a coboundary operator. If we assume in addition that $\h$ is unimodular, then
$(\HH_{pt^+},\delta^\rho)$ can be identified with the Chevalley-Eilenberg complex of Lie algebra $\h$.
Thus, the reduced space of states associated to a point (see remark \ref{Rem: H^red}) is the Chevalley-Eilenberg cohomology of $\h$:
$$\HH^\mr{red}_{pt^+}\cong H_{CE}(\h).$$
Therefore, the cohomology space
\be H_{\ad_{\hat \theta}}(Cl(\g))\cong H_{\delta_\I^\rho}(\End(\HH_{pt^+})) \cong H_{CE}(\h)\otimes (H_{CE}(\h))^* \ee
becomes non-trivial. In this case, the partition function $Z_\I^\circ$ can be understood as an identity operator acting on the Chevalley-Eilenberg cohomology $H_{CE}(\h)$.
\end{Rem}

\begin{Rem} \label{Rem: BF+B^3}
One can also view the one-dimensional version of the $BF$ theory with cosmological term \cite{CCRFM} as a special case of the one-dimensional Chern-Simons theory for $\g=\h\oplus \h^*$, where $\h$ is itself a quadratic Lie algebra, and the Lie algebra structure on $\g$ is given by
\be \theta=\frac{1}{2}F^{pqr}\eta^p \eta^q \bar\eta^r+\kappa \frac{1}{6} F^{pqr} \bar\eta^p \bar\eta^q \bar\eta^r  . \label{theta BF+B^3}\ee
Here $F^{pqr}$ are the structure constants of $\h$ (in an orthonormal basis) and the parameter $\kappa$ is the ``cosmological constant''. For Lie algebra $\g$, we automatically have $f^{abc} f^{abc}=0$, and remark \ref{Rem: H^red} applies in this case.

Let us denote $\g$ with Lie algebra structure defined by (\ref{theta BF+B^3}) by $\g_{BF,\kappa}$. Then, one-dimensional Chern-Simons theories with Lie algebras $\g_{BF,\kappa}$ and $\h$ are related, similarly to the 3-dimensional case \cite{CCRFM}. In particular, for continuum action on the circle we have
\begin{multline} \label{BF+B^3 - CS rel}
S_{\g_{BF,\kappa}}(\underbrace{\iota(\eta)+\bar\iota(\bar\eta)}_\psi,\underbrace{\bf\iota(A)+\bar\iota(\bar A)}_A)=\\
=\frac{1}{2\kappa}\left(S_\h \left(\eta+\kappa\;\bar\eta, {\bf A+\kappa\;\bar A}\right)- S_\h \left( \eta-\kappa\;\bar\eta, {\bf A-\kappa\;\bar A}\right)\right)
\end{multline}
where $\iota$ and $\bar\iota$ denote the embeddings of $\h$, $\h^*$ into $\g_{BF,\kappa}$ and on the right hand side we implicitly use the isomorphism $\h\cong \h^*$ given by the pairing on $\h$. Relation (\ref{BF+B^3 - CS rel}) implies the following relation for partition functions for triangulated circle $\Xi_n$ for Lie algebras $\g_{BF,\kappa}$ and $\h$:
\begin{multline}
Z_{\g_{BF,\kappa},\,\Xi_n}(\{\iota(\eta_k)+\bar\iota(\bar\eta_k)\}, \{\iota({\bf A}_k)+ \bar\iota({\bf \bar A}_k)\};\hbar)= \\
= Z_{\h,\, \Xi_n}(\{\eta_k + \kappa\; \bar\eta_k\}, \{{\bf A}_k+\kappa\; {\bf \bar A}_k\}; 2\kappa\, \hbar)\cdot
Z_{\h,\, \Xi_n}(\{\eta_k - \kappa\; \bar\eta_k\}, \{{\bf A}_k-\kappa\; {\bf \bar A}_k\}; -2\kappa\, \hbar)  .
\end{multline}
\end{Rem}

\begin{Rem}
Another special case of a one-dimensional Chern-Simons theory can be constructed from a Lie bialgebra $\h$. Here we set
$\g=\h\oplus \h^*$ with the canonical pairing and with Lie algebra structure on $\g$ defined by
$$\theta=\frac{1}{2}F^r_{pq}\eta^p \eta^q \bar\eta_r + \frac{1}{2} G^{qr}_p \eta^p \bar\eta_q \bar\eta_r .$$
Here $F^r_{pq}$ and $G^{qr}_p$ are structure constants of the Lie bracket and co-bracket on $\h$.
This is a one-dimensional version of the Lie bialgebra $BF$ theory, cf. \cite{Merkulov} (the underlying unimodular Lie bialgebra for continuum theory on circle is $\h\otimes \Omega^\bt(\s^1)$). It does not seem to enjoy any particular simplifications with respect to the general case  other than having a canonical complex polarization on $\g$.
\end{Rem}

\begin{Rem}
The odd third-order differential operator in variables $\tilde\psi,A, \eta_{out}, \bar\eta_{in}$ that appears in (\ref{QME with eta}) endows the algebra of functions $\Fun(\Pi\g\oplus \g\oplus\Pi\h\oplus \Pi\bar\h)$ with a structure of homotopy BV algebra in the sense of Tamarkin-Tsygan \cite{TT}. In general, the same applies  to $\Fun(\FF^\bulk_\Theta\oplus (\Pi\h)^{\times i(\Theta)}\oplus (\Pi\bar\h)^{\times i(\Theta)})$
for any 1-dimensional simplicial complex $\Theta$. If $\Theta$ has no boundary, this homotopy BV structure is strict.
\end{Rem}

\end{document}